\documentclass[nofootinbib,preprintnumbers,amsmath,amssymb,aps,prc,onecolumn, showkeys,floatfix,superscriptaddress]{revtex4-1}

\usepackage{amsmath,graphicx,color}
\usepackage{graphicx} 
\usepackage{booktabs}
\usepackage{multirow}
\usepackage{graphics}
\usepackage{hyperref}
\usepackage{bm}
\usepackage{calligra}
\usepackage[T1]{fontenc}
\usepackage{amssymb}
\usepackage{amsfonts}
\usepackage{mathrsfs}  
\usepackage{dsfont}
\usepackage{footmisc}
\usepackage[normalem]{ulem}
\usepackage{xcolor}

\graphicspath{{figures/}}

\begin{document}

\title{Causality violations in simulations of large and small heavy-ion collisions}

\author{Renata Krupczak}
\email{rkrupczak@physik.uni-bielefeld.de}
\affiliation{%
 Departamento de F\'{i}sica, Centro de Ciências Físicas e Matemáticas, Universidade Federal de Santa Catarina, Campus Universit\'{a}rio Reitor Jo\~{a}o David Ferreira Lima, Florian\'{o}polis, Brazil, 88040-900}%
 \affiliation{%
 Fakultät für Physik, Universität Bielefeld, D-33615 Bielefeld, Germany
 }%
 
\author{Tiago Nunes da Silva}
\email{t.j.nunes@ufsc.br}
\affiliation{%
 Departamento de F\'{i}sica, Centro de Ciências Físicas e Matemáticas, Universidade Federal de Santa Catarina, Campus Universit\'{a}rio Reitor Jo\~{a}o David Ferreira Lima, Florian\'{o}polis, Brazil, 88040-900}%

\author{Thiago S. Domingues}
\email{thiago.siqueira.domingues@usp.br}
\affiliation{
 Instituto de F\'isica, Universidade de S\~ao Paulo, R. do Mat\~ao, 1371, S\~ao Paulo, Brazil, 05508-090
}%

\author{Matthew Luzum}
\email{mluzum@usp.br}
\affiliation{
 Instituto de F\'isica, Universidade de S\~ao Paulo, R. do Mat\~ao, 1371, S\~ao Paulo, Brazil, 05508-090
}%

\author{Gabriel S. Denicol}
\affiliation{%
  Instituto de F\'isica, Universidade Federal Fluminense,
  Av. Milton Tavares de Souza, Niter\'oi, Brazil, 24210-346,
}%

\author{Fernando G. Gardim}
\affiliation{Instituto de Ci\^encia e Tecnologia, Universidade Federal de Alfenas, 37715-400 Po\c cos de Caldas, MG, Brazil
}

\author{Andre V. Giannini}
\affiliation{%
  Faculdade de Ciências Exatas e Tecnologia, Universidade Federal da Grande Dourados, 
  Dourados, MS, Brazil, 79804-970
}
 
\author{Mauricio N. Ferreira}
 \affiliation{%
 Department of Theoretical Physics and IFIC,  University of Valencia and CSIC, E-46100, Valencia, Spain.
}%

\author{Mauricio Hippert}
 \affiliation{%
  Illinois Center for Advanced Studies of the Universe \& Department of Physics, 
University of Illinois at Urbana-Champaign, Urbana, IL 61801-3003, USA
}

\author{Jorge Noronha}
\affiliation{%
  Illinois Center for Advanced Studies of the Universe \& Department of Physics, 
University of Illinois at Urbana-Champaign, Urbana, IL 61801-3003, USA
}%

\author{David D. Chinellato}
 \affiliation{%
  Universidade Estadual de Campinas (Unicamp), R. S\'ergio Buarque de Holanda, 777, Campinas, Brazil, 13083-859
}
  \affiliation{Stefan Meyer Institute for Subatomic Physics
of the Austrian Academy of Sciences, Wiesingerstraße 4
1010 Vienna, Austria}

\author{Jun Takahashi}
 \affiliation{%
  Universidade Estadual de Campinas (Unicamp), R. S\'ergio Buarque de Holanda, 777, Campinas, Brazil, 13083-859
}

\collaboration{The ExTrEMe Collaboration}

\date{\today}

\begin{abstract}
Heavy-ion collisions, such as Pb-Pb or p-Pb, produce extreme conditions in temperature and density that make the hadronic matter transition to a new state, called quark-gluon plasma (QGP). Simulations of heavy-ion collisions provide a way to improve our understanding of the QGP’s properties. These simulations are composed of a hybrid description that results in final observables in agreement with accelerators like LHC and RHIC. However, recent works pointed out that these hydrodynamic simulations can display acausal behavior during the evolution in certain regions, indicating a deviation from a faithful representation of the underlying QCD dynamics. To pursue a better understanding of this problem and its consequences, this work simulated two different collision systems, Pb-Pb and p-Pb at $\sqrt{s_{NN}} = 5.02$ TeV. In this context, our results show that causality violation, even though always present, typically occurs on a small part of the system, quantified by the total energy fraction residing in the acausal region. In addition, the acausal behavior can be reduced with changes in the pre-hydrodynamic factors and the definition of the bulk-viscous relaxation time.  
Since these aspects are fairly arbitrary in current simulation models, without solid guidance from the underlying theory, it is reasonable to use the disturbing presence of acausal behavior in current simulations to guide improvements towards more realistic modeling. 
While this work does not solve the acausality problem, it sheds more light on this issue and also proposes a way to solve this problem in simulations of heavy-ion collisions.
\end{abstract}

\maketitle

\section{Introduction}
\label{intro}
The study of relativistic heavy-ion collisions in accelerators such as RHIC and the LHC has revealed the production of a new phase of strongly interacting matter known as the quark-gluon plasma (QGP) \cite{Collins, Cabibbo, busza2018heavy, shuryak1978quark}. This deconfined phase of matter, formed under extreme conditions of energy and temperature, exhibits collective behavior \cite{ollitrault1992anisotropy, arsene2005quark, he2021interplaying} so that its evolution is typically described by relativistic viscous hydrodynamics \cite{israel1976nonstationary}.

Besides the QGP production in collisions of large nuclei, such as Pb-Pb or Au-Au, collisions of smaller systems, as p-Pb or d-Au \cite{aad2013measurement, nagle2018small}, have also exhibited signals of collective behavior \cite{ollitrault2009effect}, and the possibility of QGP formation in small systems, even if only for the most central collisions \cite{gardim2022smallest}, is under active debate. In fact, in proton-nucleus collisions, the QGP is produced in a very small volume and its lifetime is considerably shorter, leading to a fluid-dynamical evolution that is not close to local thermodynamic equilibrium. 

From the phenomenological point of view, the state-of-the-art method for describing the evolution of the matter formed in heavy-ion collisions involves hybrid models \cite{Petersen:2008dd}. Bayesian studies based on hybrid models have recently been vastly explored for the extraction of the hydrodynamic transport coefficients and other model parameters from experimental data \cite{moreland2020bayesian, everett2021multisystem, nijs2021bayesian}.

\begin{figure}[ht!]
     \centering
    \includegraphics[scale=0.51]{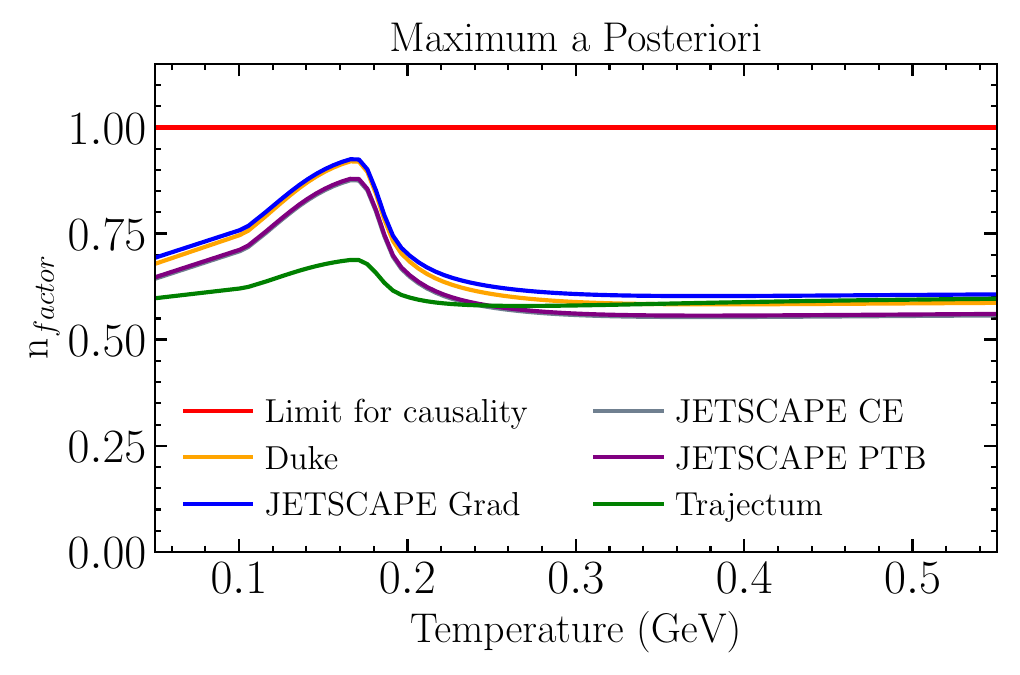}
    \includegraphics[scale=0.51]{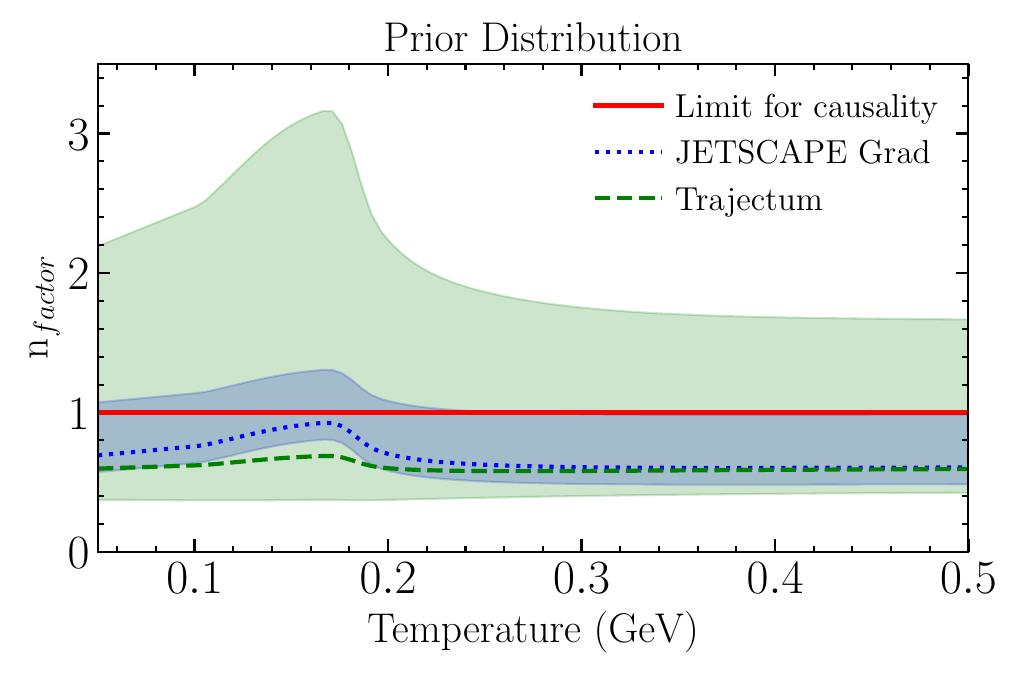}
    \caption{The linear condition for causality must be smaller than 1. In the left figure, curves corresponding to the maximum a posteriori parameter sets of several Bayesian analyses \cite{moreland2020bayesian, everett2021multisystem, nijs2021bayesian} indicate that the linearized causality condition is respected. However, in the figure on the right, we can see the prior distribution (represented by the bands around the posterior value) for JETSCAPE \cite{everett2021multisystem} and Trajectum \cite{nijs2021bayesian} violates the linear causality condition, raising the question of whether significant regions of the posterior distribution are acausal.}
    \label{fig:1}
\end{figure}

At the core of hybrid modeling lies the hydrodynamic evolution based on a second-order relativistic viscous hydrodynamics \cite{Israel:1979wp,denicol2012derivation}. In such Israel-Stewart-like theories, it has been known for a long time that  transport coefficients must satisfy the condition \cite{Hiscock:1983zz,Olson:1990rzl,Romatschke:2009im}
\begin{equation}
    n_\text{factor} \equiv c_s^2 + \frac{4}{3} \frac{\eta}{\tau_\pi (\epsilon + p)} + \frac{\zeta}{\tau_\Pi (\epsilon + p)} \leq 1,
\end{equation}
to guarantee that the characteristic velocity associated with hydrodynamic disturbances around equilibrium does not exceed the speed of light. Above, $c_s$ is the equilibrium speed of sound, $\epsilon$ and $p$ are the energy density and equilibrium pressure, respectively, while $\eta$ is the shear viscosity, $\zeta$ is the bulk viscosity, and $\tau_\pi$ and $\tau_\Pi$ are the shear and bulk relaxation times, respectively. Causality is necessary for the stability of such disturbances around equilibrium in relativity \cite{Bemfica:2020zjp,Gavassino:2021owo}. Thus, when this condition is not fulfilled, one should, in principle, expect the presence of instabilities. Therefore, for the sake of consistency, such linear constraints should not be violated in hydrodynamic simulations. Nevertheless, it is important to remark that this fundamental condition has not been imposed on prior probability distributions for the parameters in most modern Bayesian studies, as seen in Fig.~\ref{fig:1}.

Recent works have derived even more fundamental constraints to ensure that such fluid-dynamical theories respect causality in the fully general, nonlinear regime \cite{Bemfica:2019cop,bemfica2021nonlinear}. These constraints involve sufficient conditions, and also necessary conditions, for causality to be satisfied, in the form of inequalities that relate transport coefficients and the viscous currents (such as the bulk scalar $\Pi$ and information about the shear stress tensor $\pi_{\mu\nu}$) during the hydrodynamic evolution. These conditions can be applied locally to fluid elements that can then be classified as causal (if the sufficient conditions are fulfilled), acausal (if one or more of the necessary conditions are violated), or indeterminate (meaning that none of the necessary conditions were violated, but the sufficient conditions were not satisfied). These inequalities were applied in a study of different hybrid models \cite{plumberg2022causality} (see also \cite{chiu2021exploring} for a related study), which has revealed that all of the considered models violated causality at the initial stages of the hydrodynamic evolution, indicating that a problem may exist in our current modeling of heavy-ion collisions. 

In addition, Ref.\ \cite{plumberg2022causality} has shown that the pre-hydrodynamic model used in the simulation can play a crucial role in the amount of acausal fluid cells that exist in numerical simulations, both in large and small systems \cite{da2021prehydrodynamic}. Another work \cite{chiu2021exploring} has investigated the effects of different choices for the transport coefficients on the necessary conditions, and a slight change in flow observables was observed.

Building on the previous research above, this work aims to analyze the causes of acausality in collisions of large and small nuclei and compare the results found in such systems. This paper is organized as follows. Section \ref{sec.model} describes the computational framework and the model parameters used in this work. Then,  Section \ref{sec.results} discusses the results and presents different comparisons. First, in subsection \ref{subsec.causality}, we compare large collision systems (Pb-Pb) with small systems (p-Pb), set in the same way in energy and basic parameters. After that, in subsection \ref{subsec.FS}, we investigate how the changes in the pre-hydrodynamic stage can reduce the acausality found in the simulations. In the third subsection \ref{subsec.eq} we analyze and discuss the violated conditions. Finally, in Section \ref{sec.conclusions} we present our conclusions.

\section{Numerical Setup}
\label{sec.model}
We have simulated p-Pb and Pb-Pb collisions at center-of-mass energy $\sqrt{s_{NN}} = 5.02$ TeV employing a numerical chain comprised of 
\begin{itemize}
    \item $\mathrm{T}_{\mathrm{R}} \mathrm{ENTo}$ \cite{moreland2015alternative}, to generate the system's initial energy density;
    \item Free-streaming of the initial $\mathrm{T}_{\mathrm{R}} \mathrm{ENTo}$ profile using \cite{liu2015pre};
    \item MUSIC code \cite{schenke20103+} for the viscous hydrodynamical evolution;
    \item frzout code for particlization based on the Cooper-Frye equation \cite{cooper1974single};
    \item UrQMD \cite{bleicher1999relativistic} as a hadronic cascade.
\end{itemize}
The simulations were performed for $2000$ minimum bias events for each system. The parameters for $\mathrm{T}_{\mathrm{R}} \mathrm{ENTo}$ and free-streaming, and QGP transport coefficients were obtained from a simultaneous Bayesian analysis of both colliding systems performed in \cite{moreland2020bayesian}. Table IV of Ref.\  \cite{moreland2020bayesian} lists all parameters used here.

For the initial analysis of causality violations in the nonlinear regime, the simulations were performed until the end of the hydrodynamic evolution at a freeze-out hypersurface defined by a constant temperature of 151 MeV \cite{moreland2020bayesian}. For this study, the hydrodynamical variables were saved in intervals of $0.5$ fm/c for Pb-Pb collisions and in intervals of $0.25$ fm/c for p-Pb collisions, given its shorter evolution. For calculating the final state observables, the particlization, and hadronic cascade stages were also simulated following the hydrodynamic evolution.
The equation of state used in the hydrodynamic evolution is constructed by matching the HotQCD EoS \cite{bazavov2019chiral} with a hadronic resonance gas that has the same matter content as UrQMD\footnote{\url{https://github.com/j-f-paquet/eos_maker}}. We do not expect that using different equations of state, such as \cite{Alba:2017hhe}, would significantly change our results.

Local transport coefficients and viscous currents at fixed time steps were calculated in each fluid cell during the hydrodynamic evolution, allowing for the analysis of causality violations.  The extracted quantities were used for evaluating the necessary and sufficient conditions for causality obtained in \cite{bemfica2021nonlinear}. Each cell was then classified accordingly.

In Section~\ref{sec.results} we present a study of the dependence of the causality violation on some of the parameters in our simulations. In this case, only the selected parameter was varied, while all other ones were kept fixed.

\section{Results}
\label{sec.results}
\subsection{Causality violation in Pb-Pb and p-Pb collisions}
\label{subsec.causality}

\begin{figure*}
    \centering
    \begin{minipage}[!]{0.24\linewidth}
    \includegraphics[scale=0.27]{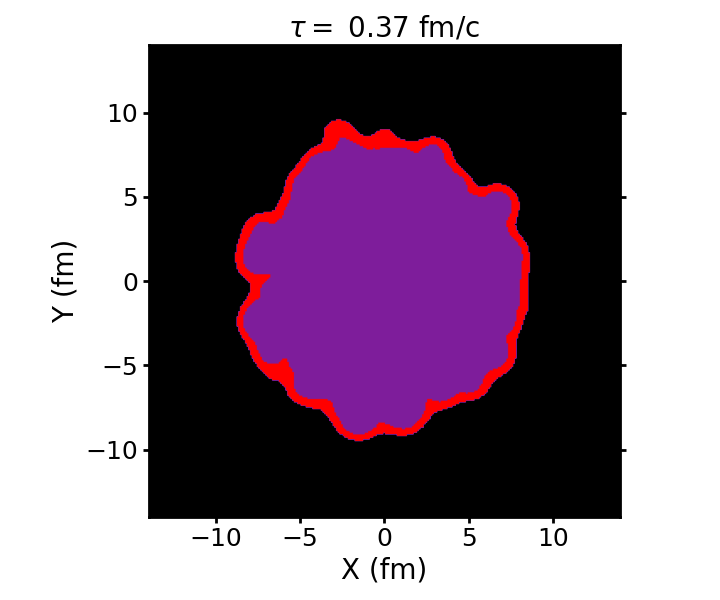}
    \end{minipage}
    \begin{minipage}[!]{0.24\linewidth}
    \includegraphics[scale=0.27]{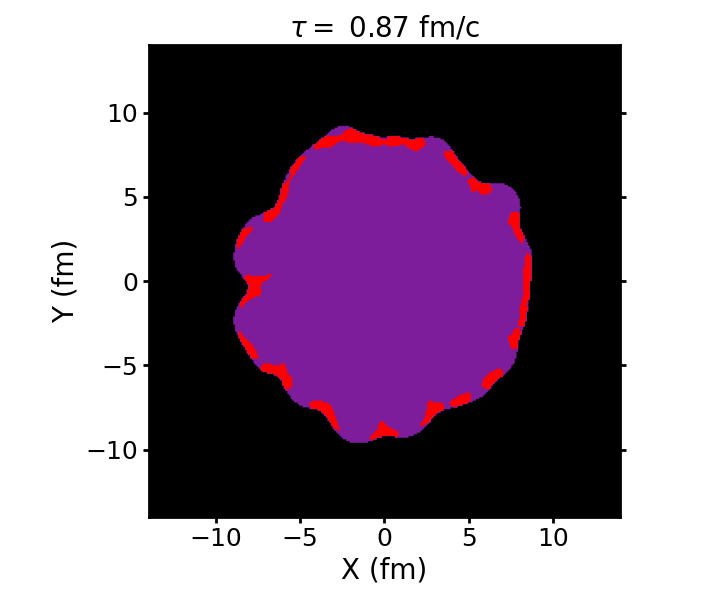}
    \end{minipage}
    \begin{minipage}[!]{0.24\linewidth}
    \includegraphics[scale=0.27]{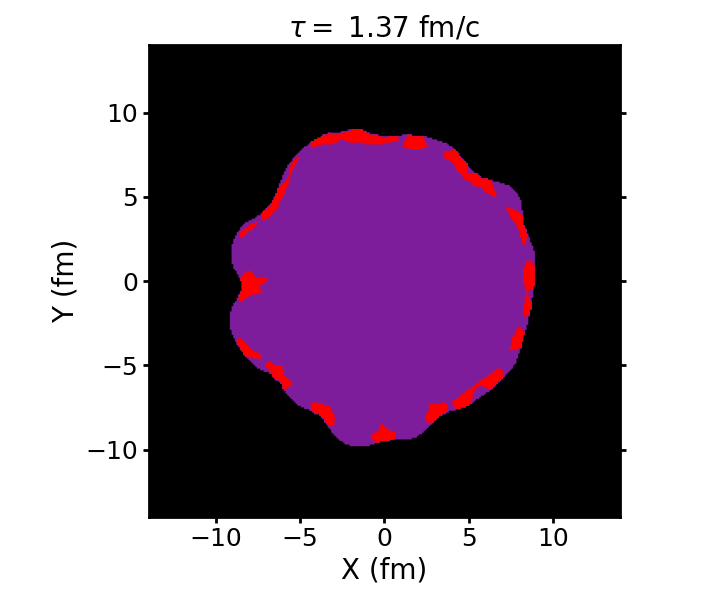}
    \end{minipage}
    \begin{minipage}[!]{0.24\linewidth}
    \includegraphics[scale=0.27]{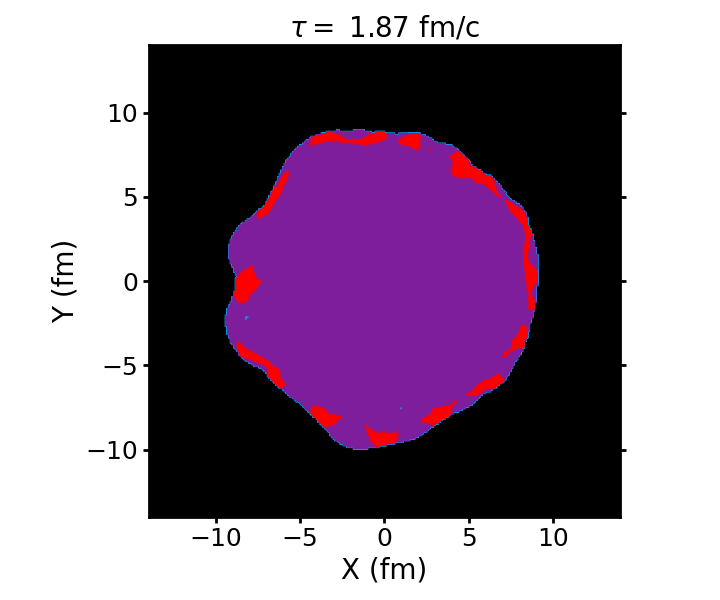}
    \end{minipage}
    \begin{minipage}[!]{0.24\linewidth}
    \includegraphics[scale=0.27]{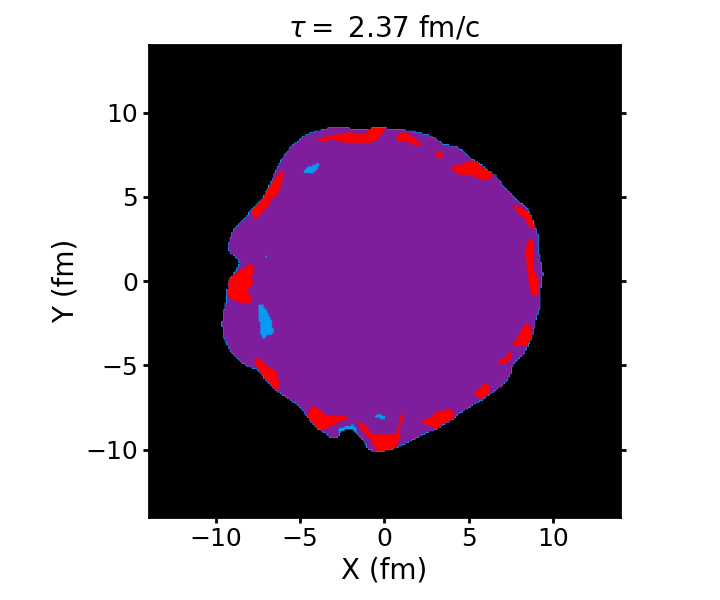}
    \end{minipage}
    \begin{minipage}[!]{0.24\linewidth}
    \includegraphics[scale=0.27]{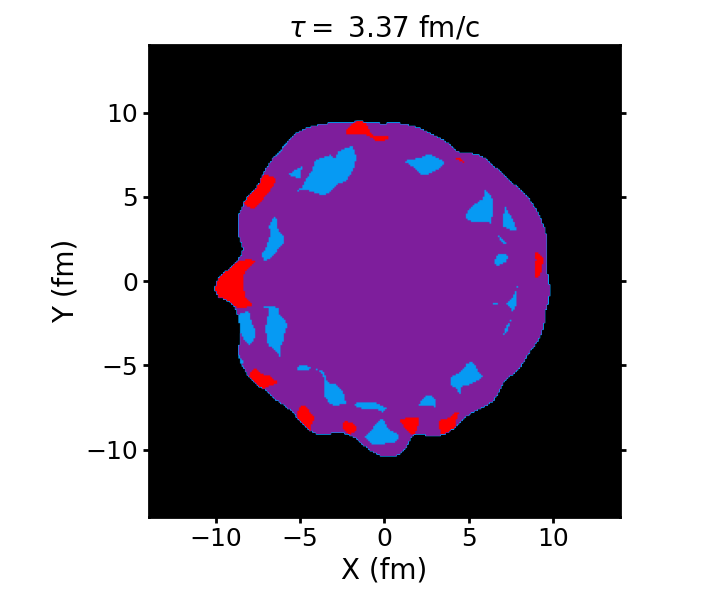}
    \end{minipage}
    \begin{minipage}[!]{0.24\linewidth}
    \includegraphics[scale=0.27]{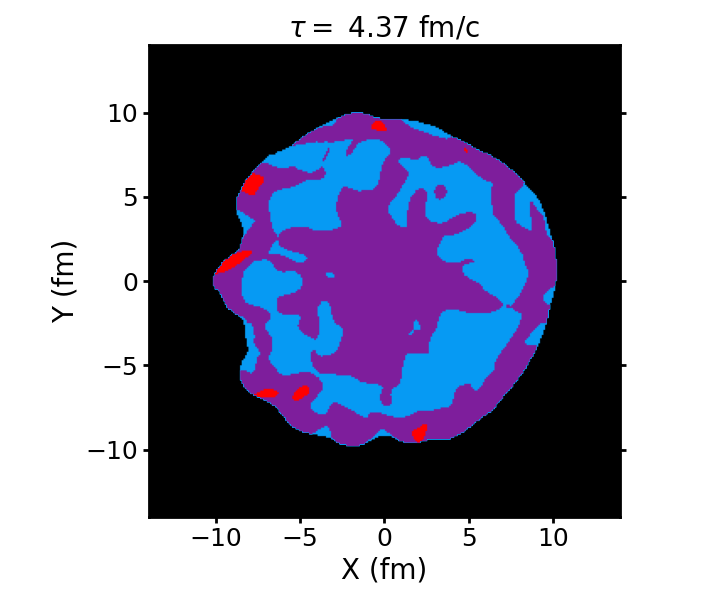}
    \end{minipage}
    \begin{minipage}[!]{0.24\linewidth}
    \includegraphics[scale=0.27]{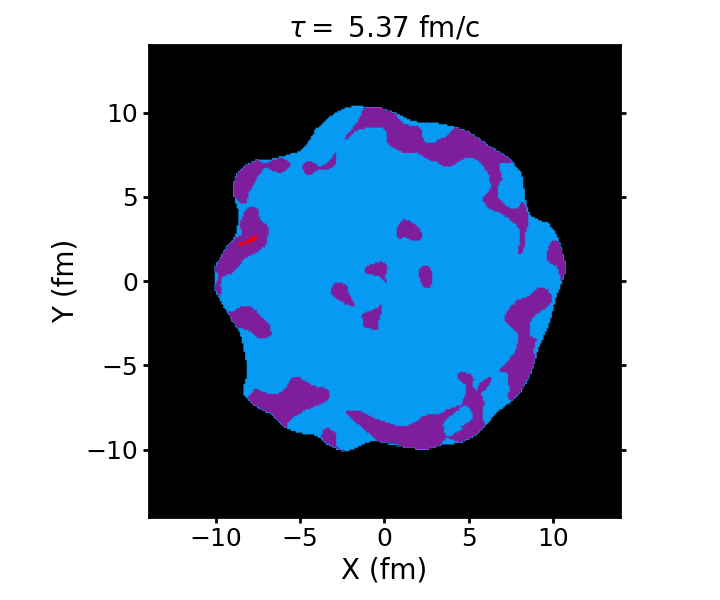}
    \end{minipage}
    \begin{minipage}[!]{0.24\linewidth}
    \includegraphics[scale=0.27]{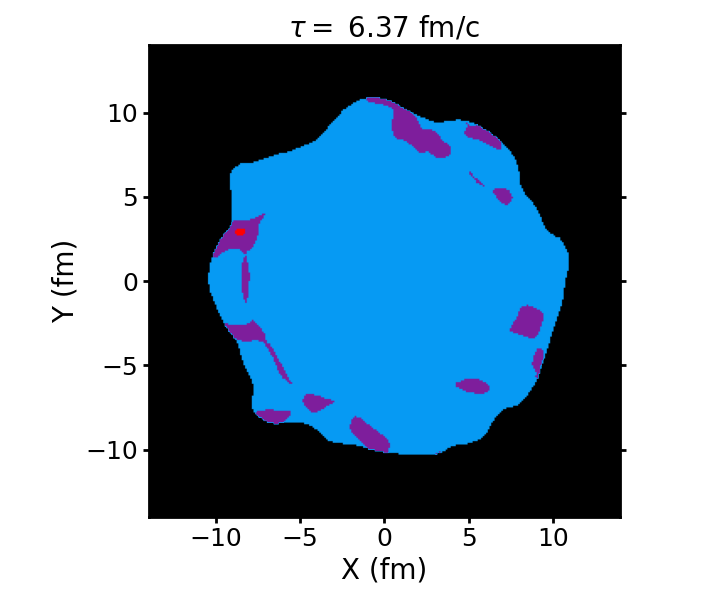}
    \end{minipage}
    \begin{minipage}[!]{0.24\linewidth}
    \includegraphics[scale=0.27]{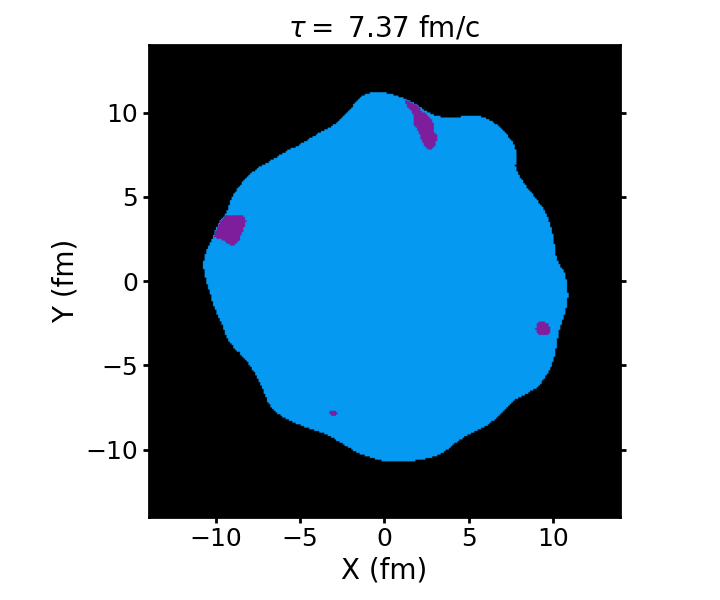}
    \end{minipage}
    \begin{minipage}[!]{0.24\linewidth}
    \includegraphics[scale=0.27]{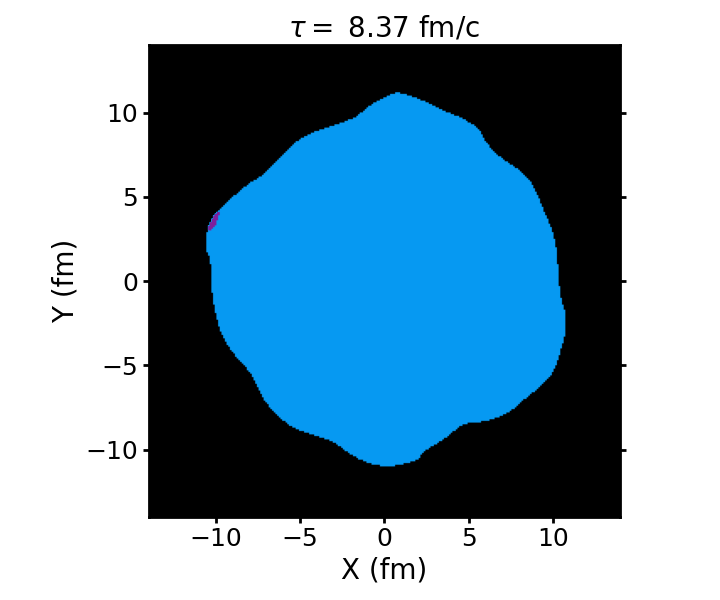}
    \end{minipage}
    \begin{minipage}[!]{0.24\linewidth}
    \includegraphics[scale=0.27]{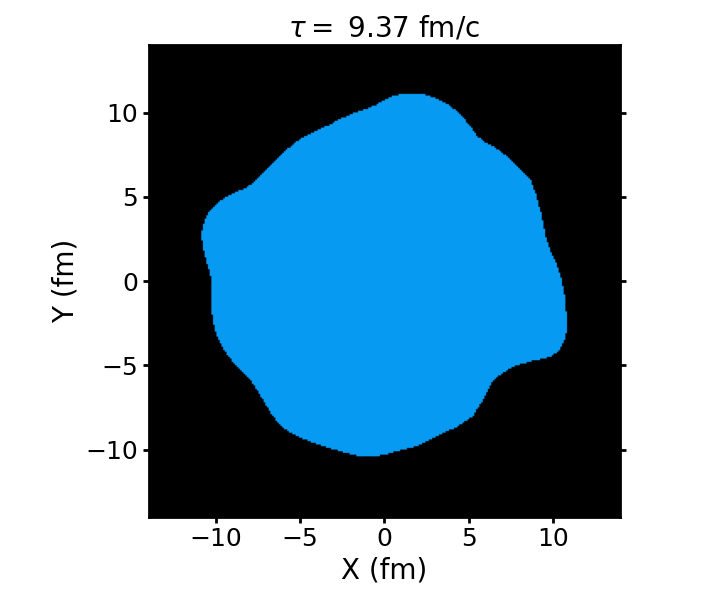}
    \end{minipage}
    \caption{Causality analysis in a Pb-Pb event with centrality $0-5\%$, free-streaming time of $0.37$ fm/c, end time of hydrodynamics equals $13.21$ fm/c, and the freezeout temperature equal to $151$ MeV. The classification scheme is causal (blue), indeterminate (purple), and acausal (red), as discussed in the previous section.}
    \label{fig2}
\end{figure*}

\begin{figure*}
    \centering
    \includegraphics[scale=0.51]{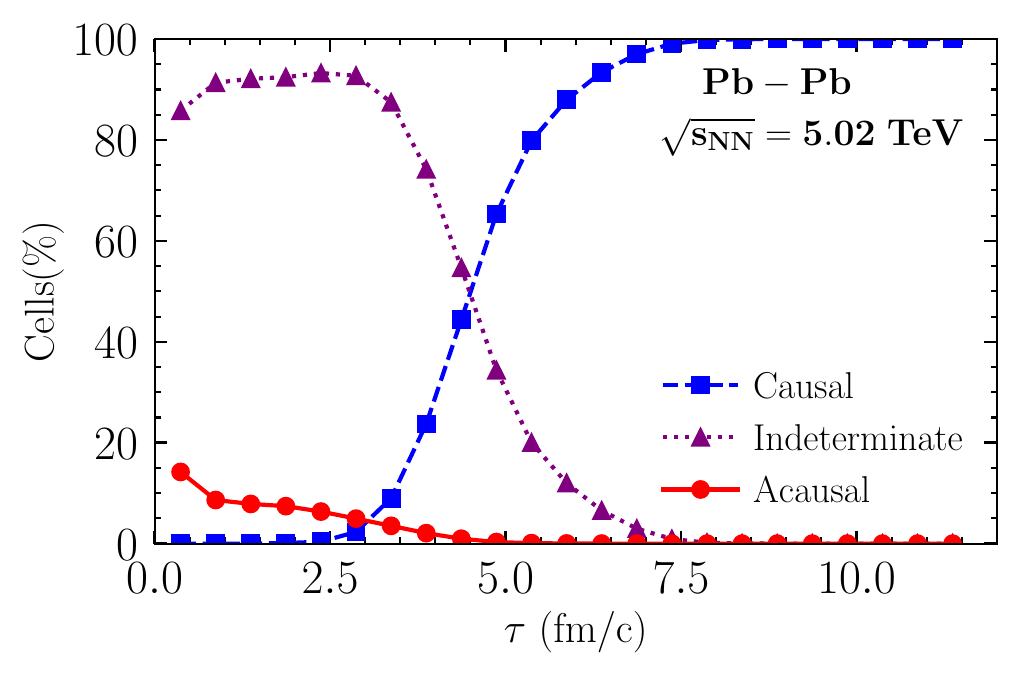}
    \caption{The mean percentage of causal, acausal, and indeterminate cells versus the hydrodynamic time for events in the $0-5\%$ centrality class for Pb-Pb collisions. The line between the points is drawn only to guide the eye. The first point in time, $0.37$ fm/c, represents the end of the free-streaming stage and the beginning of hydrodynamics. In this setting, we have less than $20\%$ of the cells violating causality at the beginning of the hydrodynamic evolution.}
    \label{fig3}
\end{figure*}

\begin{figure}
    \centering
    \includegraphics[scale=0.51]{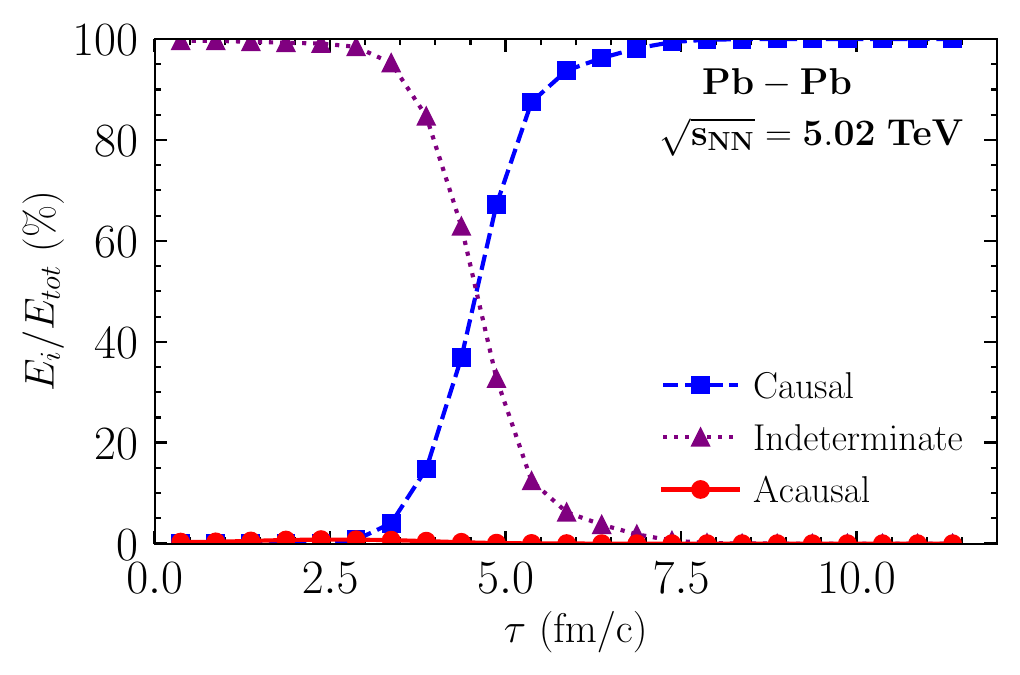}
    \includegraphics[scale=0.51]{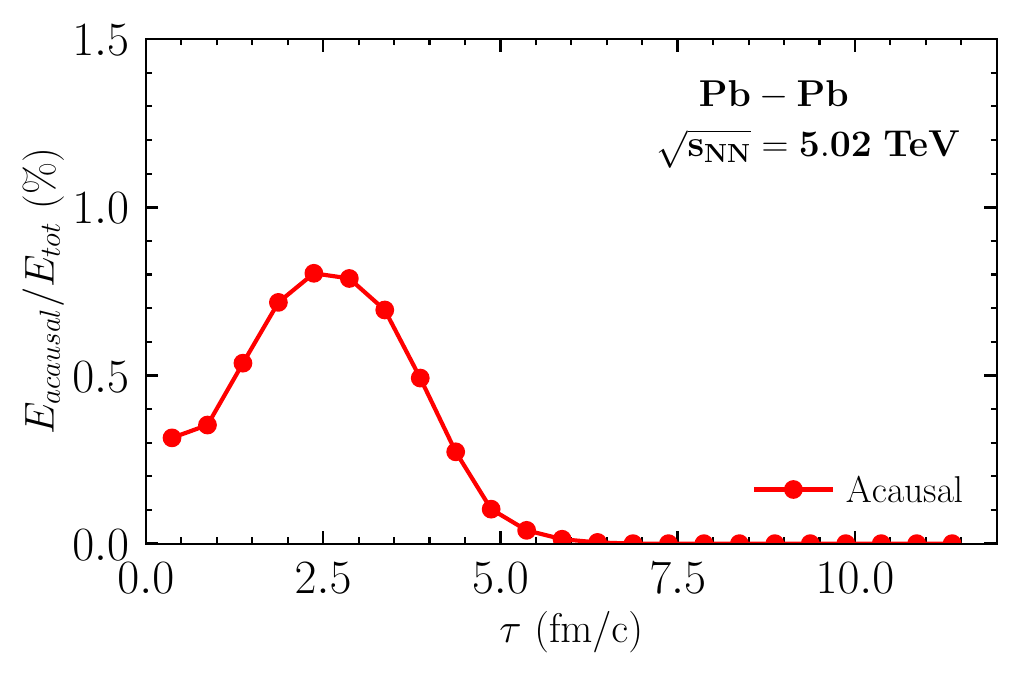}
    \caption{(Left panel) The mean percentage of energy for different properties ($i$): causal, acausal, and indeterminate versus the hydrodynamic time for Pb-Pb events in the $0-5\%$ centrality class. (Right panel) Focus on the energy percentage of acausal cells versus the hydrodynamic time for events in the $0-5\%$  centrality class. Though the number of acausal cells is significant (Fig. \ref{fig3}), the fraction of energy in such cells is only a small part of the energy of the whole system (less than $1\%$).}
    \label{fig4}
\end{figure}

We show in Fig.\ \ref{fig2} how the percentage of causal (blue), acausal (red), and indeterminate (purple) cells change during the hydrodynamic stage of a Pb-Pb collision at $5.02$ TeV, for an event in the $0-5\%$ centrality class, using the numerical chain described above. This outcome reproduces the same qualitative results as those found in \cite{plumberg2022causality}. In this work, the pre-hydrodynamic model is used in every scenario, as it significantly reduces the occurrence of acausal cells at the beginning of the hydrodynamic phase. We note that our events were simulated at higher collisional energy and with different free-streaming time ($\tau_{fs} = 0.37$ fm/c) in comparison to \cite{plumberg2022causality}, and we observe less than $20\%$ of acausal cells at the start of hydrodynamics, as shown in Fig.\ \ref{fig3}, which is a slightly lower value than the results showed in \cite{plumberg2022causality} for Pb-Pb events at $2.76$ TeV. This acausal behavior is mostly present at the edge of the system and is reduced to zero along the hydrodynamic evolution, which lasts around $12$ fm/c for this centrality class. 

\begin{figure}
    \centering
    \includegraphics[scale=0.51]{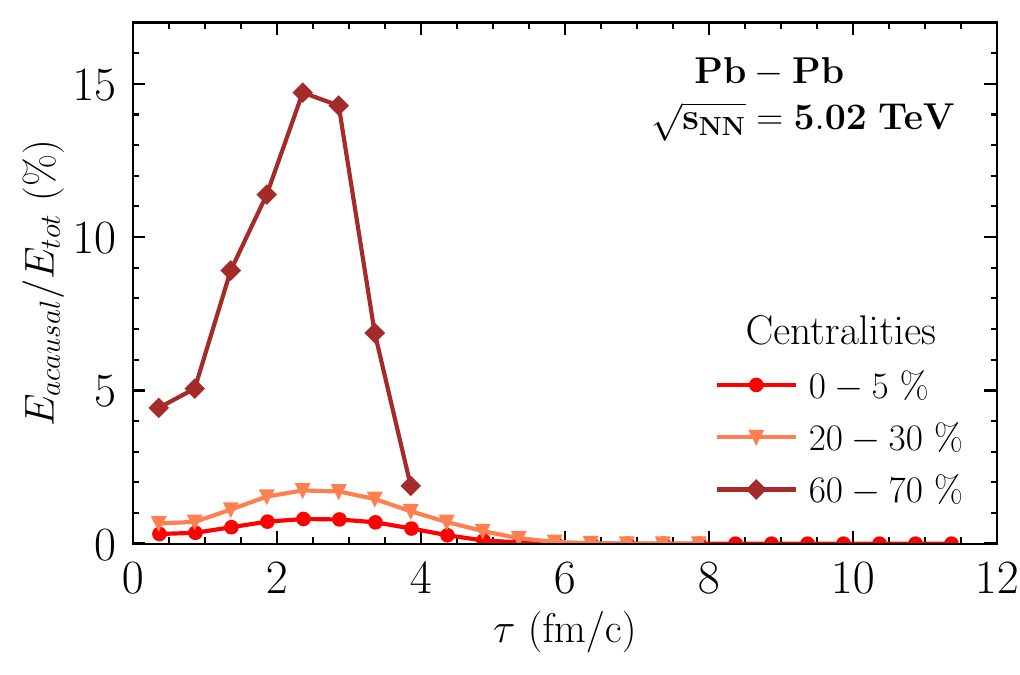}
    \caption{The mean percentage of acausal energy in Pb-Pb events at different centralities.}
    \label{fig.5}
\end{figure}

\begin{figure*}
    \centering
    \begin{minipage}[!]{0.24\linewidth}
    \includegraphics[scale=0.27]{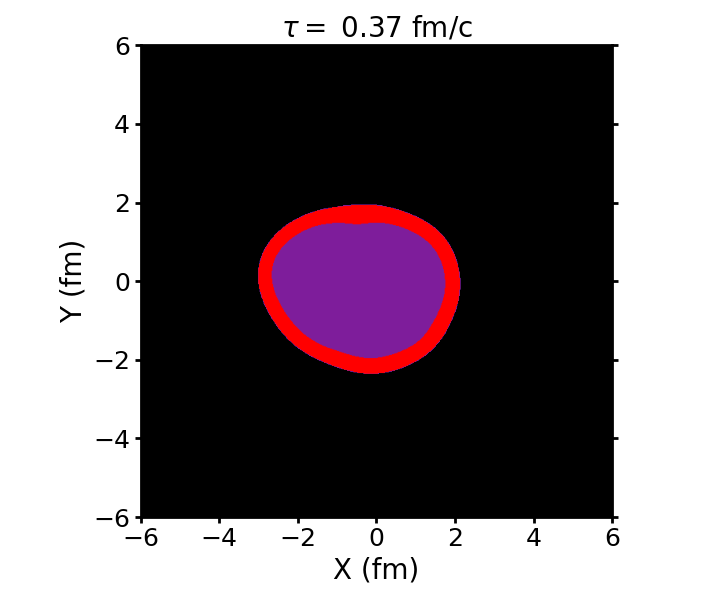}
    \end{minipage}
    \begin{minipage}[!]{0.24\linewidth}
    \includegraphics[scale=0.27]{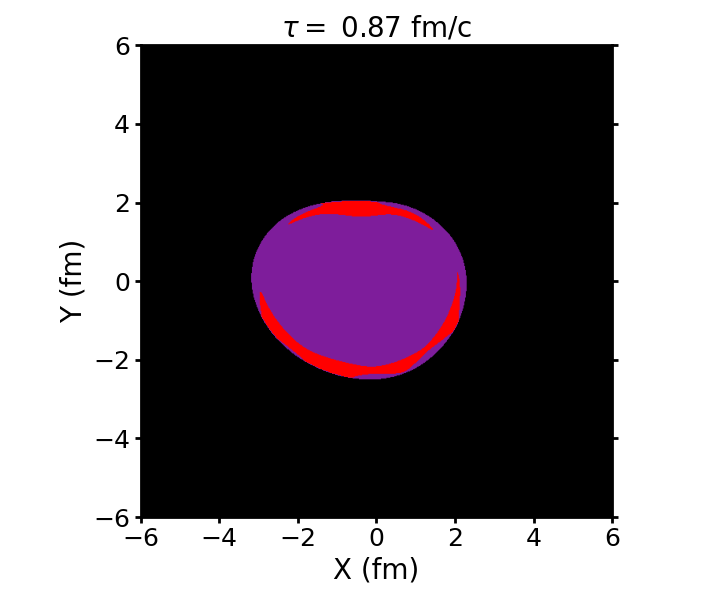}
    \end{minipage}
    \begin{minipage}[!]{0.24\linewidth}
    \includegraphics[scale=0.27]{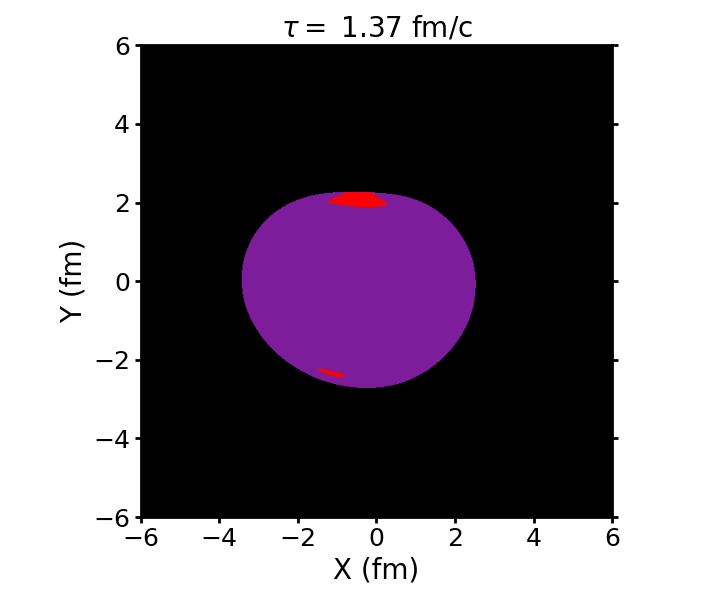}
    \end{minipage}
    \begin{minipage}[!]{0.24\linewidth}
    \includegraphics[scale=0.27]{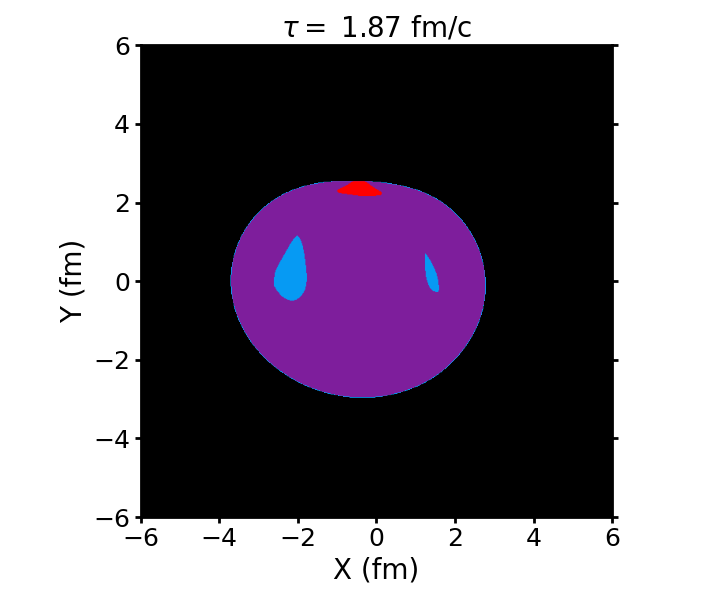}
    \end{minipage}
    \begin{minipage}[!]{0.24\linewidth}
    \includegraphics[scale=0.27]{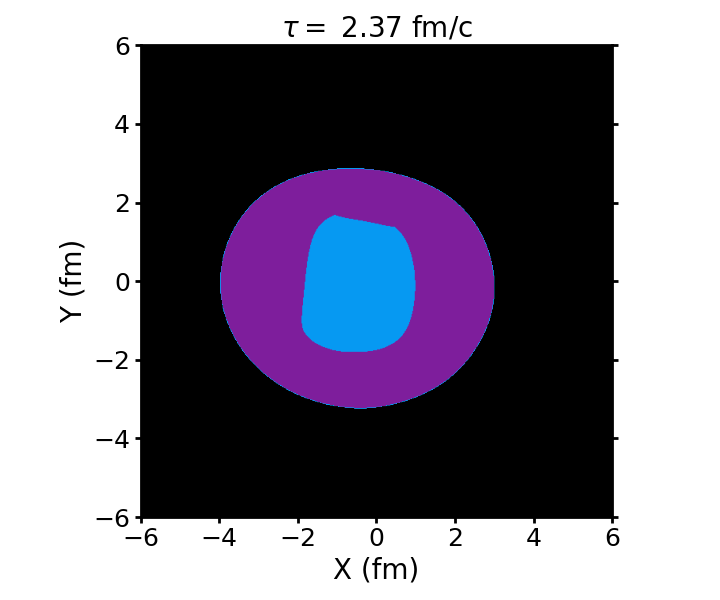}
    \end{minipage}
    \begin{minipage}[!]{0.24\linewidth}
    \includegraphics[scale=0.27]{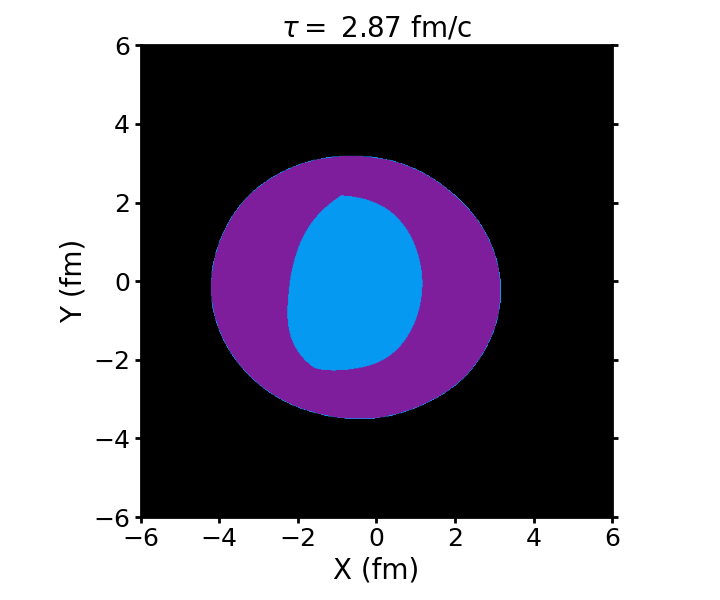}
    \end{minipage}
    \begin{minipage}[!]{0.24\linewidth}
    \includegraphics[scale=0.27]{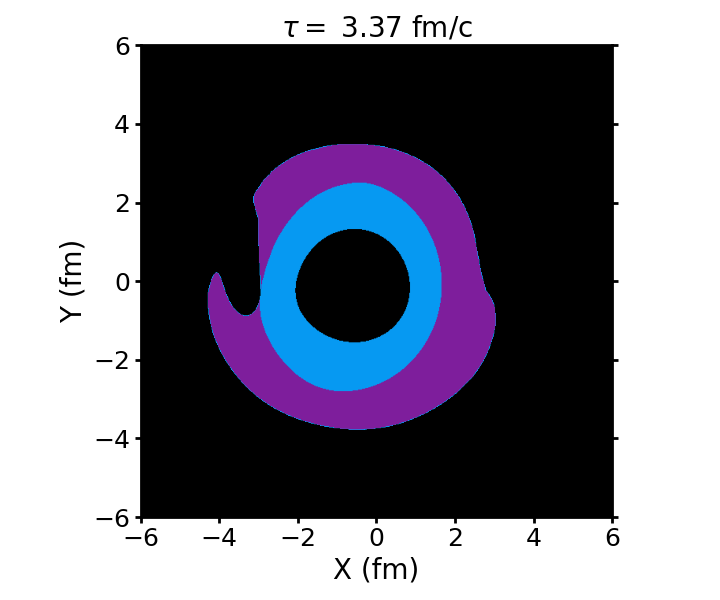}
    \end{minipage}
    \begin{minipage}[!]{0.24\linewidth}
    \includegraphics[scale=0.27]{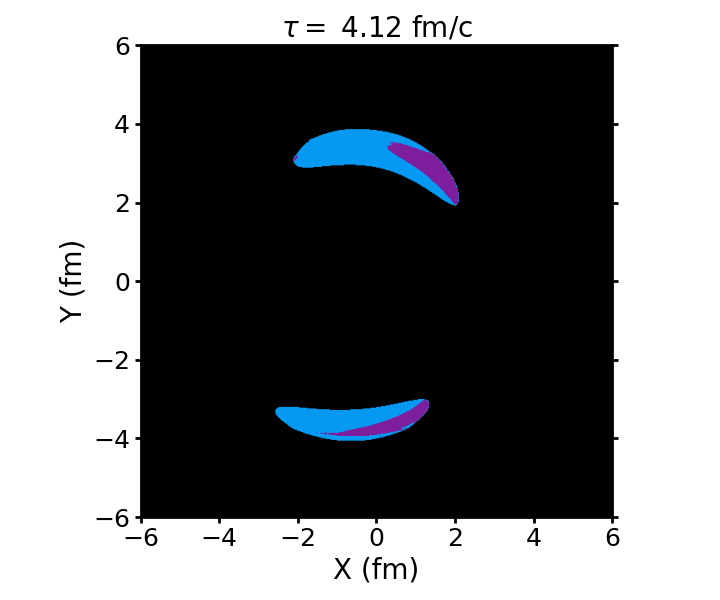}
    \end{minipage}
    \caption{Analysis of causality in a p-Pb event with $0-5\%$ centrality, free-streaming time equal to $0.37$ fm/c, and final hydrodynamic time equals $4.47$ fm/c. The color system is the same as before. One can see that this leads to results similar to those in Fig.\ \ref{fig2}. The hole shown in the last panels appears due to hadronization. Here, the freezeout temperature is $151$ MeV. Lower temperatures are in the hadron gas phase, which has not been analyzed in this study.}
    \label{fig6}
\end{figure*}

\begin{figure}
    \centering
    \begin{minipage}[!]{1\linewidth}
    \includegraphics[scale=0.51]{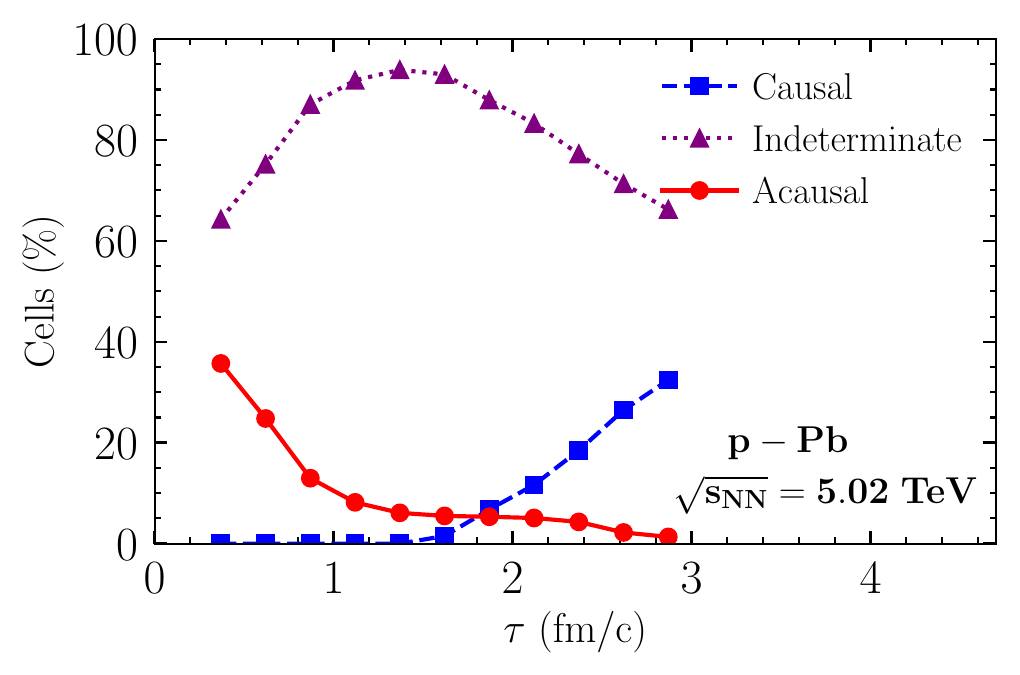}
    \end{minipage}
    \caption{The mean percentage of causal, acausal, and indeterminate cells in p-Pb events in the $0-5\%$ centrality class versus the time of the hydrodynamic evolution. In this setting, we have around $36\%$ of acausal cells at the beginning of hydrodynamics. The color pattern follows the same scheme used before: red is for acausal cells, blue is for causal cells, and purple is for indeterminate cells.}
    \label{fig7}
\end{figure}

\begin{figure}
    \centering
    \includegraphics[scale=0.51]{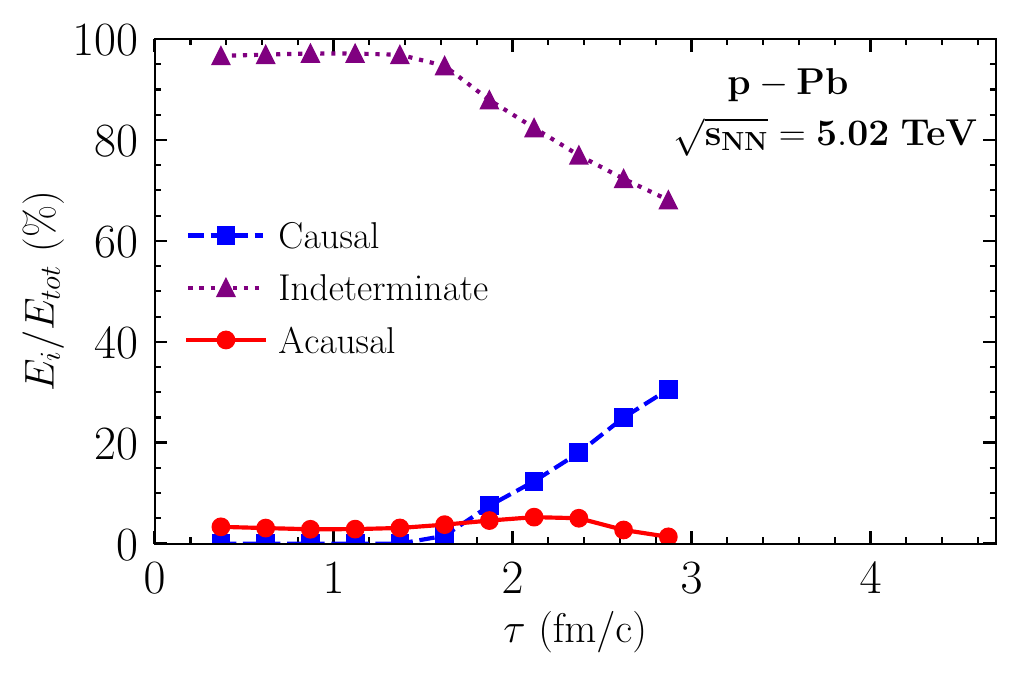}
    \includegraphics[scale=0.51]{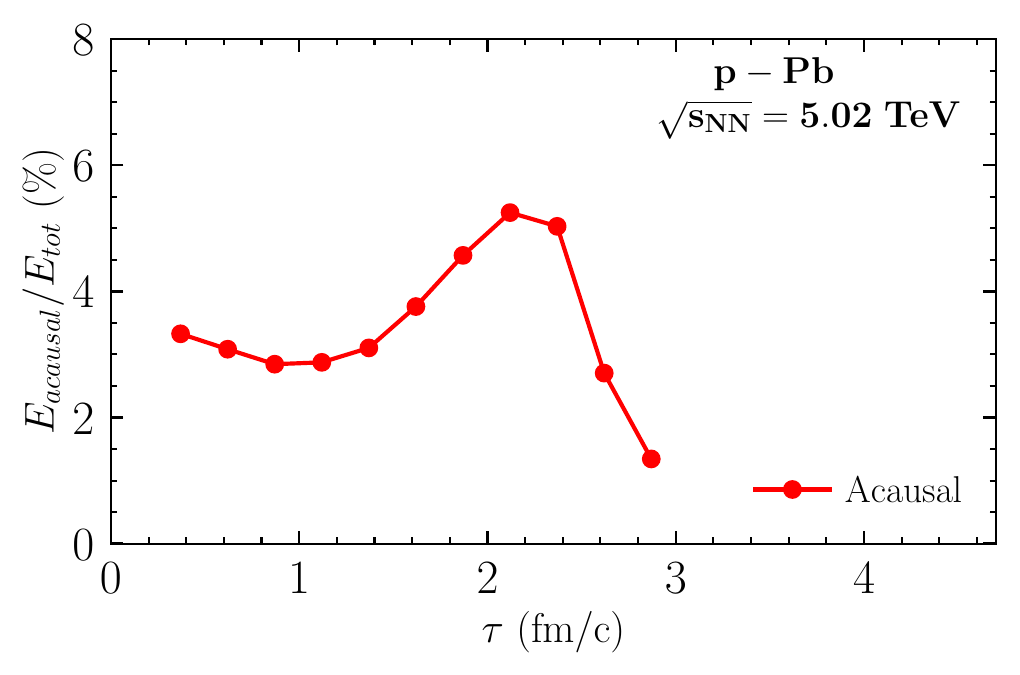}
    \caption{(Left panel) The mean percentage of energy for different properties: causal, acausal, and indeterminate for central events ($0-5\%$) in p-Pb collision versus the hydrodynamic time. (Right panel) Focus on the energy percentage in acausal cells versus the hydrodynamic time. The line between the points is drawn only to guide the eye.}
    \label{fig8}
\end{figure}

In \cite{plumberg2022causality}, causality violation was quantified by counting the fraction of non-causal cells in the fluid --- that is, the transverse area covered by fluid in an acausal regime compared to the total area of the fluid. In the worst scenarios, roughly $75\%$ of the cells exhibited causality violations at the beginning of hydrodynamics. Since for our initial conditions most of the violations occur at the edges\footnote{We note that causality violation can also happen in the inner parts of the fluid if different initial conditions are used, as shown in \cite{plumberg2022causality}. The initial conditions used here, where the violations are mostly concentrated at the edges, provide the best-case scenario when it comes to causality violations.}, where the fluid is less dense, we propose to quantify causality violations by weighting each cell within the fluid by its energy content, effectively counting the percentage of energy contained in acausal cells in the system. As shown in Fig.~\ref{fig4}, obtained from an ensemble of events in the $0-5\%$ centrality class, the roughly $20\%$ of initial cells in Pb-Pb collisions with causality violation carry only approximately $0.3\%$ of the energy.

A visualization of the centrality dependence of the acausal energy fraction evolution is presented in Fig.\ \ref{fig.5}, where averages between events in a given centrality class show that less central collisions have a higher percentage of energy in acausal cells. This indicates that causality violation becomes a severe problem in hydrodynamic simulations performed in the ultraperipheral regime \cite{Zhao:2022ayk}.   

The same analysis is presented in Figs.\ \ref{fig6}, \ref{fig7}, \ref{fig8} for a smaller system, namely p-Pb collisions, in the same centrality class as the Pb-Pb analyses. Again, for our initial conditions, causality violations are concentrated at the edge of the system, and they decrease during the hydrodynamic evolution. It is interesting to note that, since the hydrodynamic evolution is shorter for this system, in some events, there is not enough time for the entire system to become causal. This illustrates how causality violations can be important to further constrain the hydrodynamic description of small systems. In comparison to the larger Pb-Pb system, we observe that p-Pb exhibits larger violations of causality, which might be associated with more significant deviations from equilibrium \cite{niemi2014large,Noronha-Hostler:2015coa}, and a larger amount of fluctuations \cite{tumasyan2023strange}. The smaller system also has the same centrality dependence as the Pb-Pb system, with more significant violations in more peripheral events, similar to what is shown in Fig.\ \ref{fig.5}. Further work is needed to investigate how these results depend on the initial state model. Here, we used $\mathrm{T}_{\mathrm{R}} \mathrm{ENTo}$ initial conditions, which are smoother than other models, such as IP-Glasma \cite{Gale:2012rq}. 

In the following, we explore how the results above can be affected by the different parameters used in our simulations.

\subsection{Variation of Free-streaming Parameters}
\label{subsec.FS}
We first investigate how free-streaming parameters can affect causality violation in the hydrodynamics phase. For example, one expects that the energy-momentum tensor resulting from the pre-equilibrium stage still possesses large gradients that would lead to causality being violated at the beginning of hydrodynamics. In addition, it is known that the instantaneous switch from the conformal regime in most pre-equilibrium models to the non-conformal QCD equation of state in hydrodynamics generates an artificially large bulk pressure at the beginning of hydrodynamics \cite{da2021prehydrodynamic}. Thus, one might wonder whether this artifact also contributes to causality violation.

\begin{figure}
    \centering
    \includegraphics[scale=0.51]{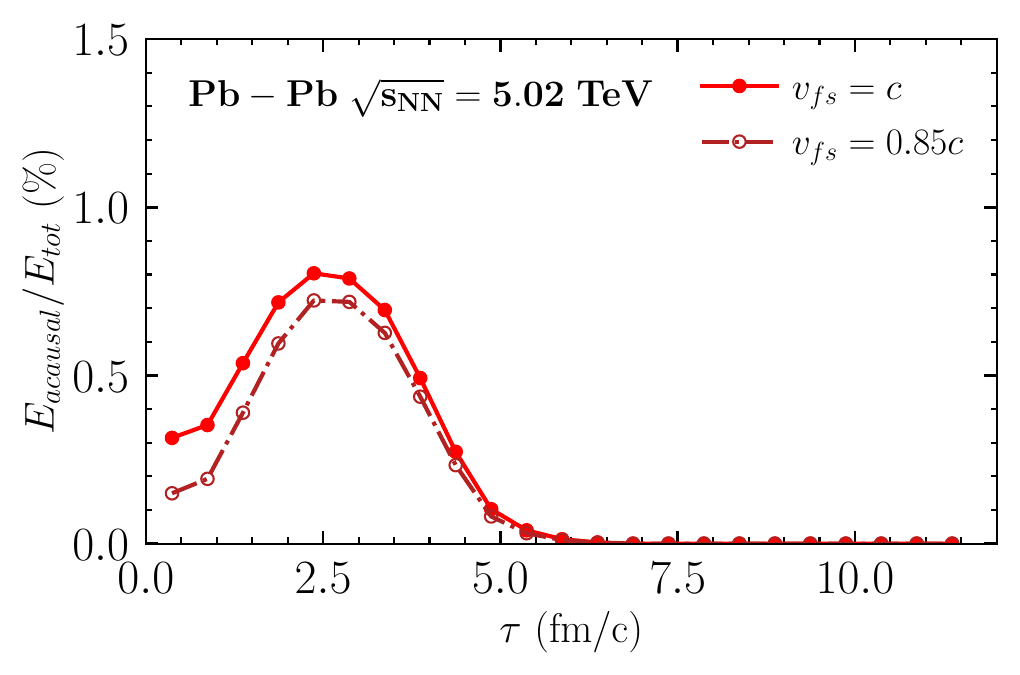}
    \includegraphics[scale=0.51]{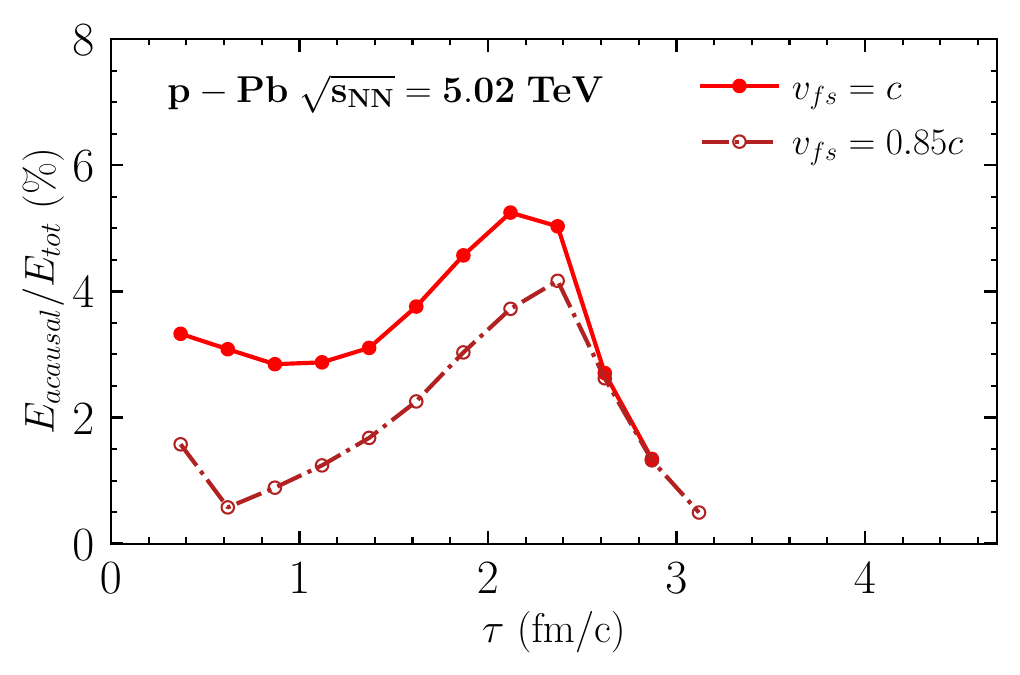}
    \caption{The mean percentage of acausal energy for different velocities in the free-streaming phase. In the left, we show results for Pb-Pb collisions and, in the right, we show results for p-Pb collisions. Both sets of results are shown in the $0-5\%$  centrality class. Using a velocity of $0.85$c in the free-streaming stage significantly reduces the causality violation in both systems.}
    \label{fig9}
\end{figure}

\begin{figure}
    \centering
    \includegraphics[scale=0.51]{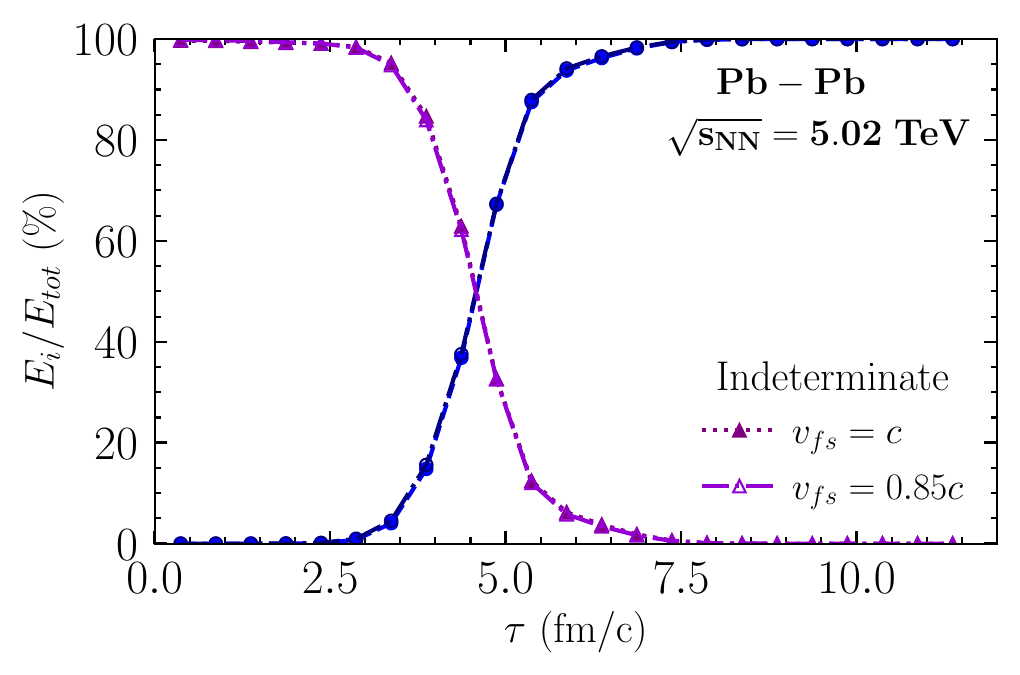}
    \includegraphics[scale=0.51]{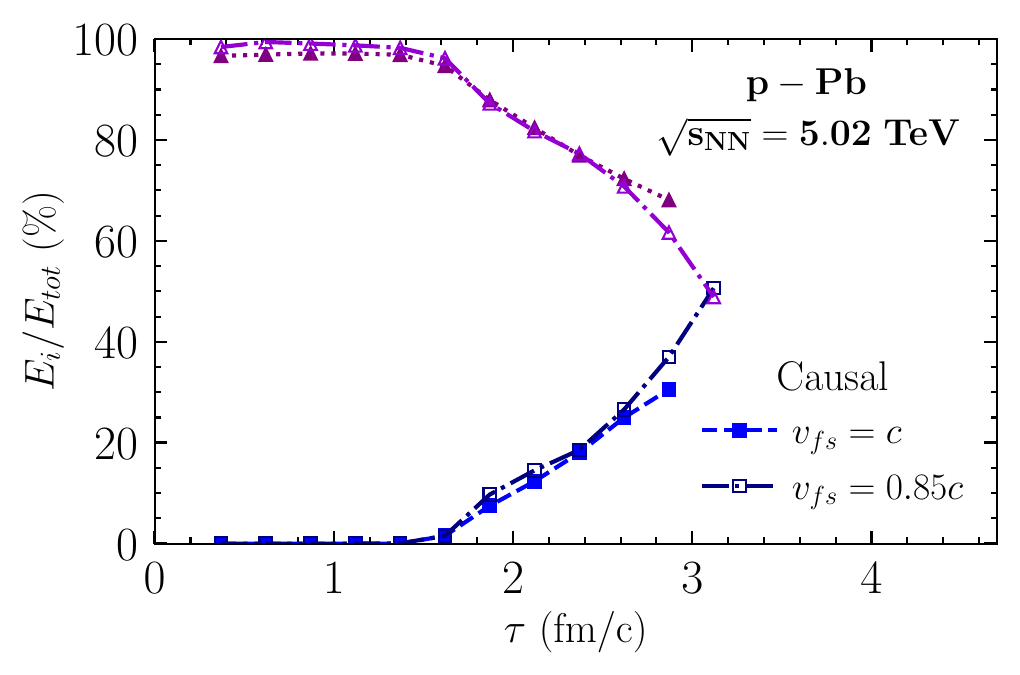}
    \caption{Percentage of energy for different properties: causal (shades of blue) and indeterminate (shades of purple) for different velocities in the free-streaming phase. On the left is the mean for Pb-Pb events, and on the right, we show p-Pb events, both for $0-5\%$ centrality. The line is added only to guide the eye. The percentage of energy in causal cells does not change.}
    \label{fig10}
\end{figure}

By default, free-streaming assumes purely transverse propagation at the speed of light. In previous work, it has been shown \cite{da2023prehydrodynamic} that a free-streaming stage with subluminal velocity effectively breaks conformal invariance and can be used to better approximate the equation of state of matter in the pre-equilibrium stage to that of QCD. We have thus compared results from our original simulations presented above with results from simulations in which the free-streaming velocity is set to $v_{fs}=0.85$c. As shown in Fig.\ \ref{fig9}, for both Pb-Pb and p-Pb, the $0.85$c curve exhibits less violation of causality, with a reduction of approximately $50\%$ of the total energy in acausal cells in the system at the first time step. We note, however, that the energy content in acausal cells becomes indeterminate as the causal cells remain unaffected, which can be seen in Fig.\ \ref{fig10}. We obtained similar outcomes for different centralities when comparing these two different velocities. We conclude that reducing the free-streaming velocity reduces the initial acausal behavior. This suggests that bulk pressure is a relevant factor but not the only parameter that controls causality violation. 

\begin{figure}
    \centering
    \includegraphics[scale=0.51]{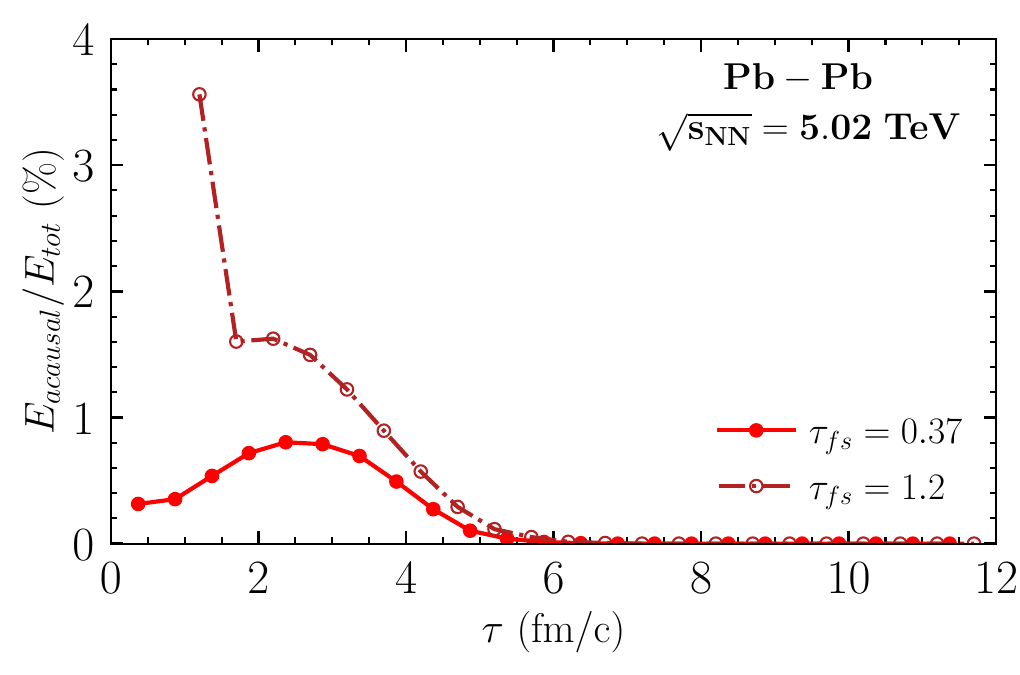}
    \includegraphics[scale=0.51]{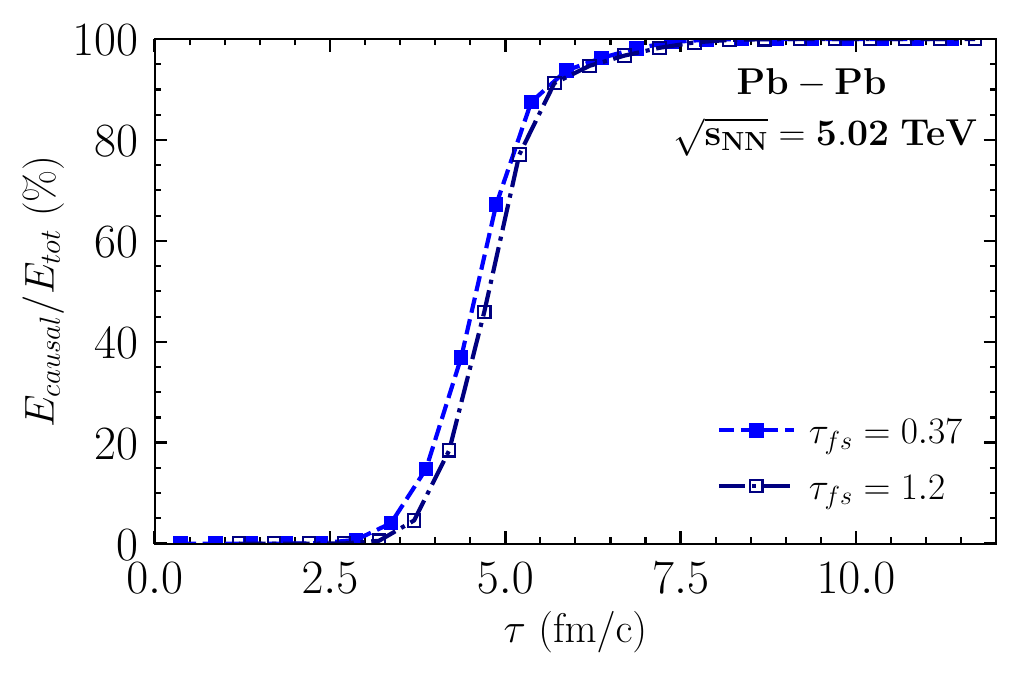}
    \includegraphics[scale=0.51]{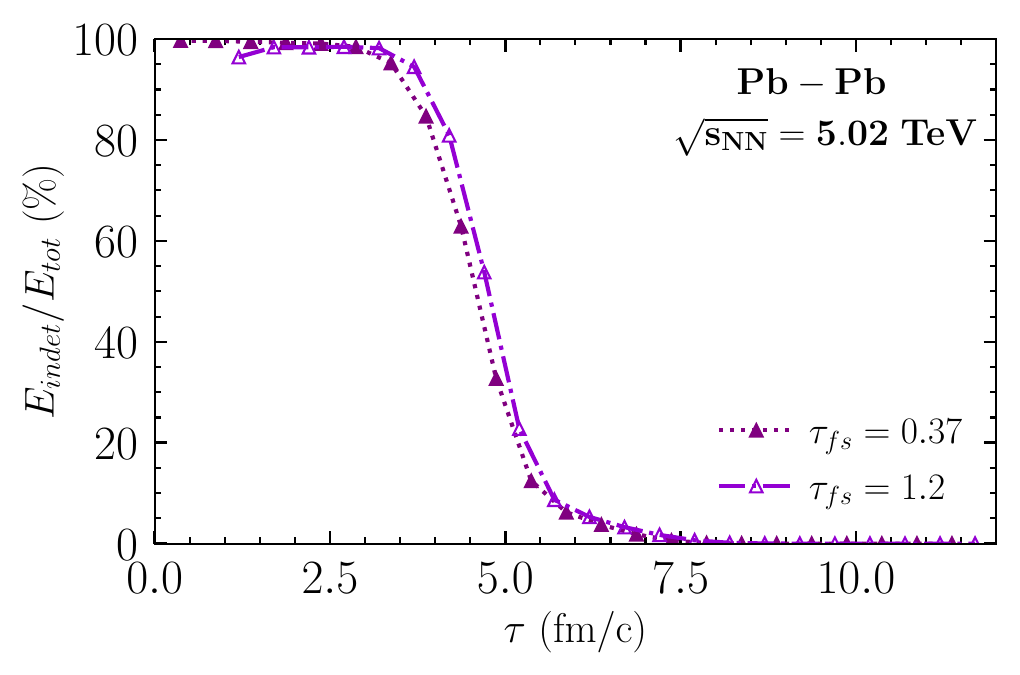}
    \caption{Mean percentage of energy for different properties: causal, acausal, and indeterminate for Pb-Pb collisions at $0-5\%$ centrality, with different free-streaming times and free-streaming velocity fixed to be the speed of light. The lines are shown only to guide the eye. Shortening the free-streaming stage reduces the initial acausality and produces a small difference in the causal energy.}
    \label{fig11}
\end{figure}

We have also explored the dependence of causality violation on the duration of the free-streaming stage. Namely, we simulate new events with $\tau_\text{fs} = 1.2$ fm/c. Results can be found in Fig.\ \ref{fig11} for Pb-Pb events, which show that longer free-streaming times lead to more causality violations. This analysis is not repeated for p-Pb due to the smaller lifetime of this system.

\subsection{Analysis of the Causality Conditions}
\label{subsec.eq}
To better understand what causes the violations, we systematically examined the violation of the nonlinear causality conditions. We investigate which conditions are typically violated when causality cannot be established for a given fluid cell. Following the notation of \cite{bemfica2021nonlinear} (we refer the reader to that work for the full expressions for the causality conditions), we find that indeterminate cells violate sufficient conditions $5b$, $5e$, $5g$, and/or $5h$. Notably, we have also found that acausal cells consistently violate the necessary condition $4f$ and the same sufficient conditions $5b$, $5e$, $5g$, and/or $5h$. These findings are identical for both Pb-Pb and p-Pb systems.

Note that acausal cells only violate one necessary equation. If this violation did not occur, the cell would be automatically converted into an indeterminate cell. The equation $4f$ in question reads:
\begin{equation}
    \begin{aligned} \epsilon+ P+\Pi+\Lambda_d &-\frac{1}{2 \tau_\pi}\left(2 \eta+\lambda_{\pi \Pi} \Pi\right) -\frac{\tau_{\pi \pi}}{2 \tau_\pi} \Lambda_d -\frac{1}{6 \tau_\pi}\left[2 \eta+\lambda_{\pi \Pi} \Pi+\left(6 \delta_{\pi \pi}-\tau_{\pi \pi}\right) \Lambda_d\right] \\ & -\frac{\zeta+\delta_{\Pi \Pi} \Pi+\lambda_{\Pi \pi} \Lambda_d}{\tau_{\Pi}}-\left(\varepsilon+P+\Pi+\Lambda_d\right) c_s^2 \geq 0~.
    \end{aligned}
\end{equation}
Above, $\Lambda_d$ is one of the eigenvalues of $\pi_{\mu\nu}$ (see \cite{bemfica2021nonlinear}), and $\tau_{\pi\pi}$, $\lambda_{\pi\Pi}$, $\delta_{\pi\pi}$, $\delta_{\Pi\Pi}$, $\lambda_{\Pi\pi}$ are second-order transport coefficients \cite{denicol2012derivation}. 

\begin{figure}
    \centering
    \includegraphics[scale=0.51]{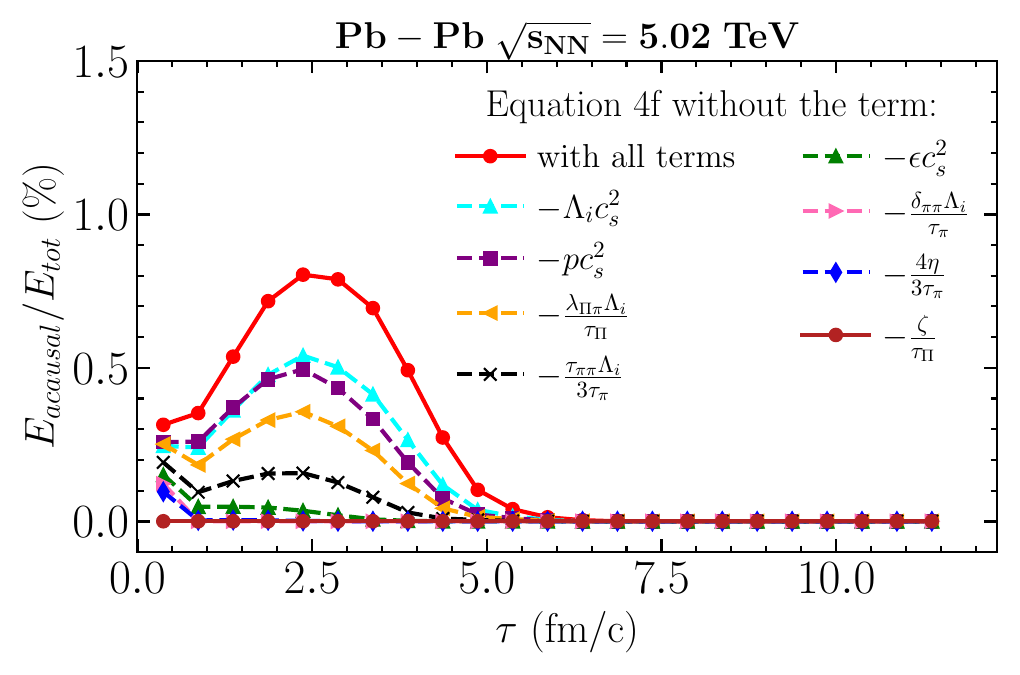}
    \includegraphics[scale=0.51]{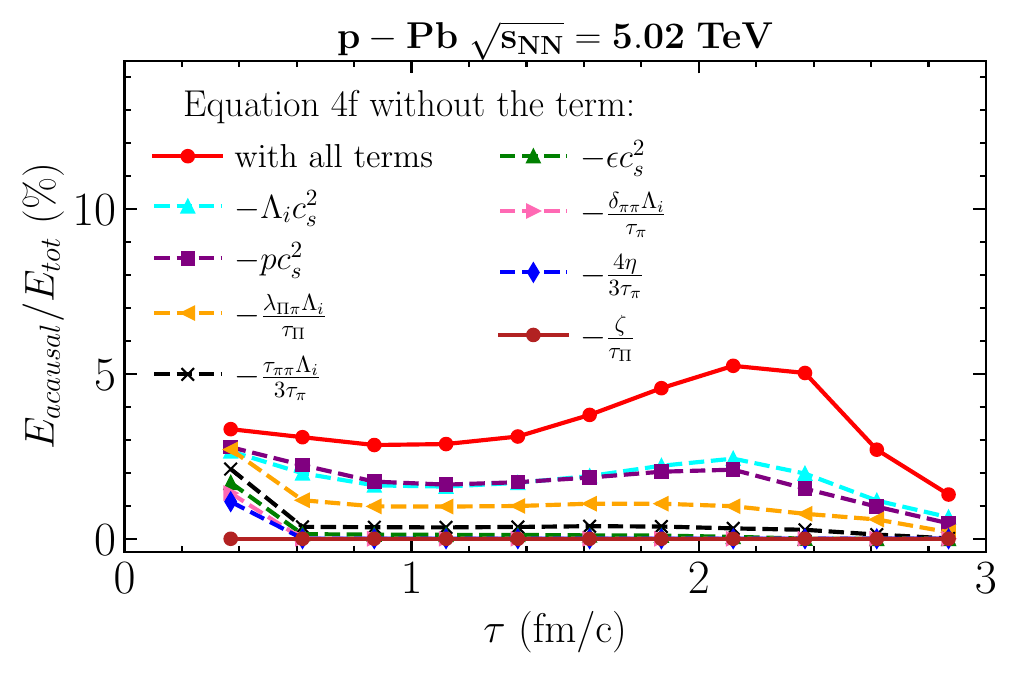}
    \caption{Mean percentage of causal energy for events with $0-5\%$ centrality for Pb-Pb (left) and p-Pb (right). The light red circles indicate the acausality using the standard analysis. The other curves are the results when omitting a term in equation $4f$. The dark red circles give the best result, where the acausality is canceled during the hydrodynamic phase.}
    \label{fig12}
\end{figure}

To understand which coefficients are responsible for causality violations, we systematically turn off each term in the equation above, while keeping the remaining terms, and perform a new causality analysis. We have found a reduction in acausality when specific terms are absent. Specifically, acausality is eliminated when the equation does not include the term $-\zeta/\tau_\Pi$. The results of this reduction of acausality are shown in Fig. \ref{fig12} for both Pb-Pb and p-Pb systems.

This indicates that these violations are closely related to the bulk viscosity parameter $\zeta$ and the bulk relaxation time $\tau_\Pi$. While it would be unphysical to eliminate this term, we note that the definition of $\tau_\Pi$ used in the hydrodynamic simulations possesses an overall constant, $b_\Pi$, that was obtained from a toy-model kinetic calculation \cite{denicol2014transport, denicol2012derivation}, and its influence on final state observables has not been thoroughly studied so far. We thus proceeded to study the dependence of the causality violations on this parameter $b_\Pi$, shown below
\begin{equation}
    \tau_\Pi = b_\Pi \frac{\zeta}{(1/3 - c_s^2)^2(\epsilon + P)}~.
\end{equation}
\begin{figure}
    \centering
    \includegraphics[scale=0.51]{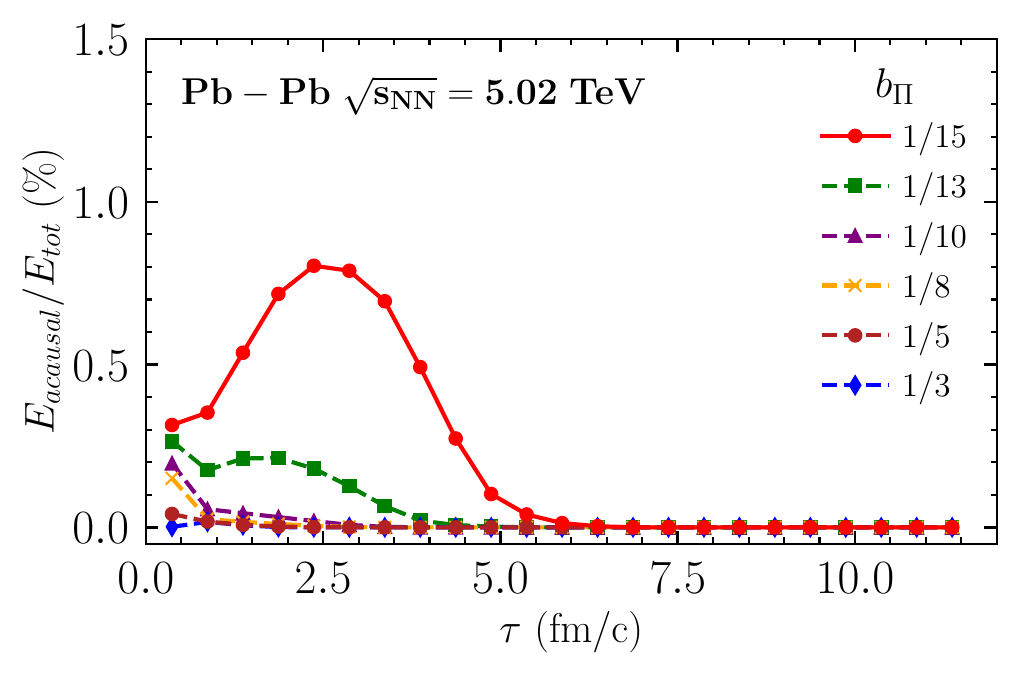}
    \includegraphics[scale=0.51]{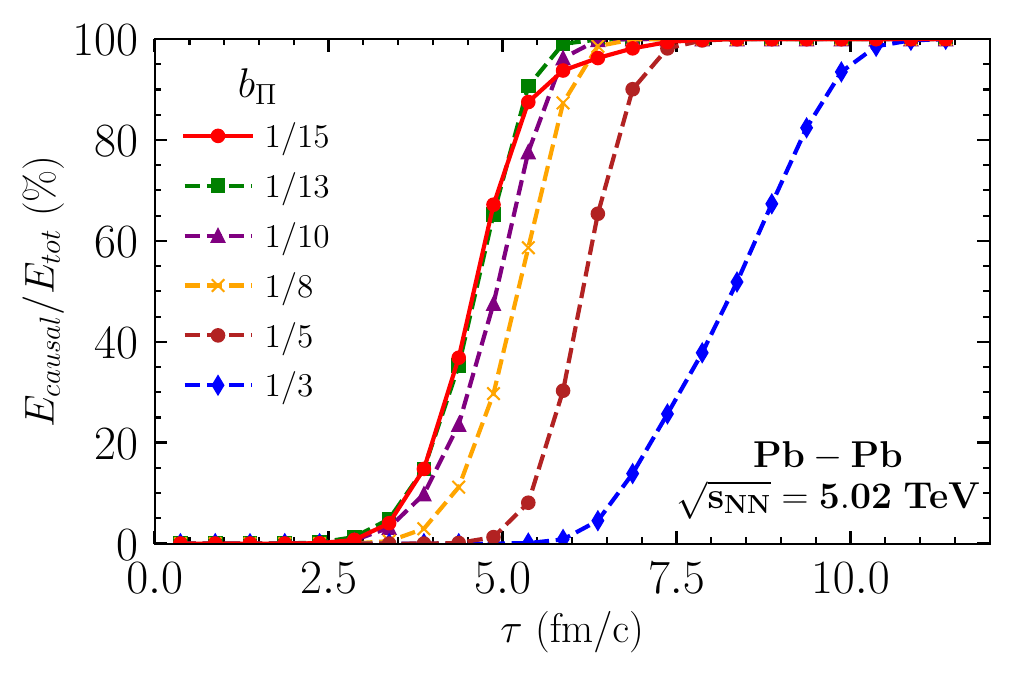}
    \includegraphics[scale=0.51]{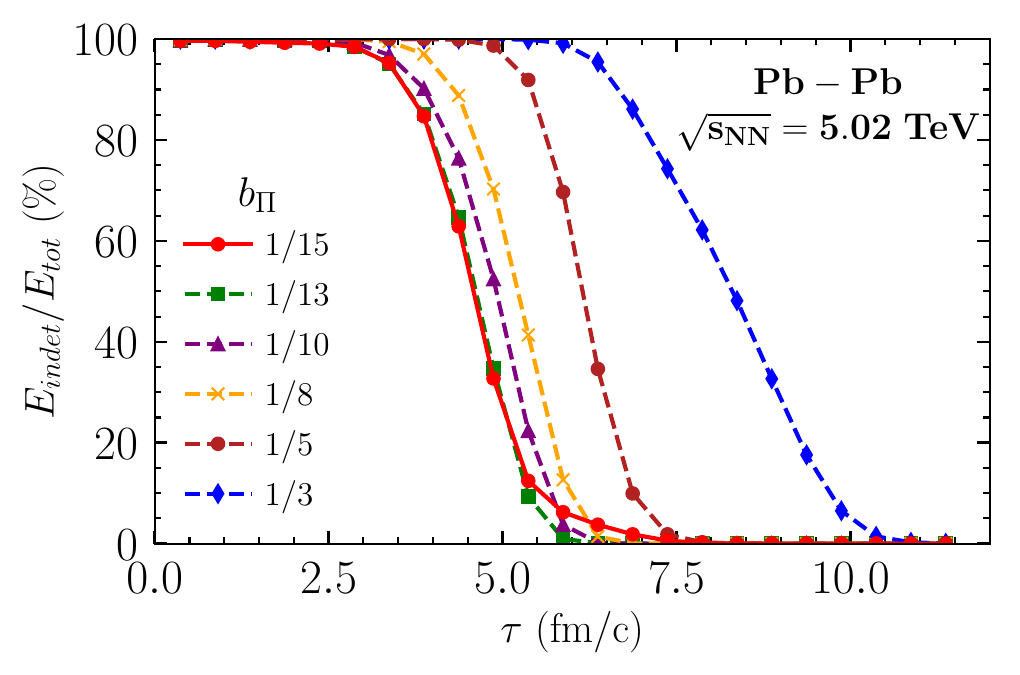}
    \caption{Different values for the parameter $b_\Pi$ and how it modifies the amount of causal, acausal, and indeterminate energy for Pb-Pb collisions. Although larger $b_\Pi$ reduces the initial acausality, it slows the conversion from acausal/indeterminate to causal cells.}
    \label{fig13}
\end{figure}

\begin{figure}
    \centering
    \includegraphics[scale=0.51]{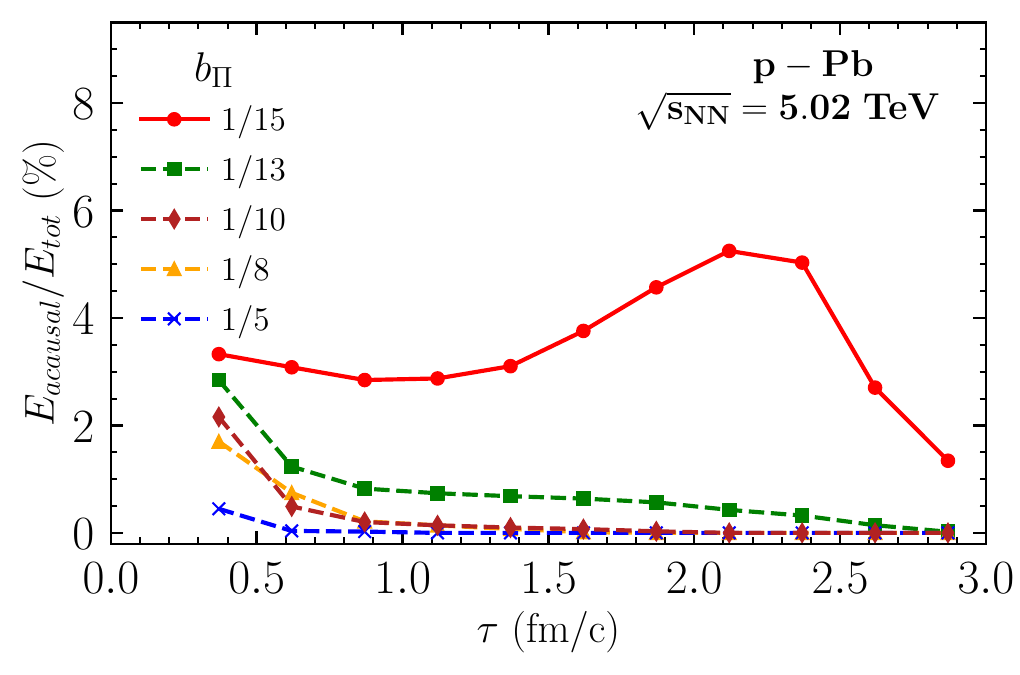}
    \includegraphics[scale=0.51]{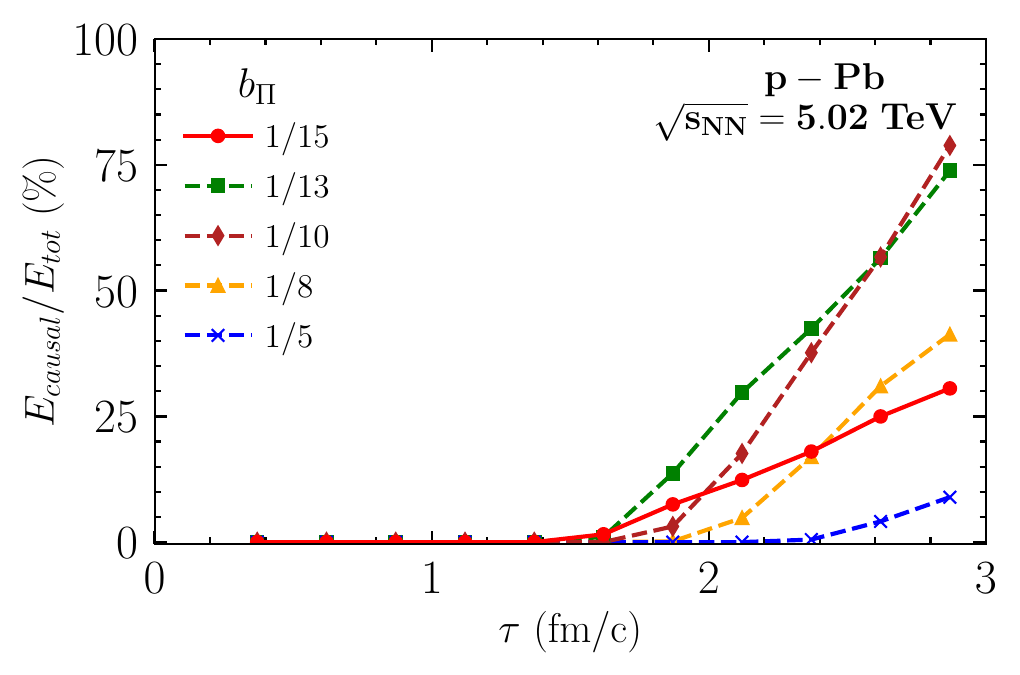}
    \includegraphics[scale=0.51]{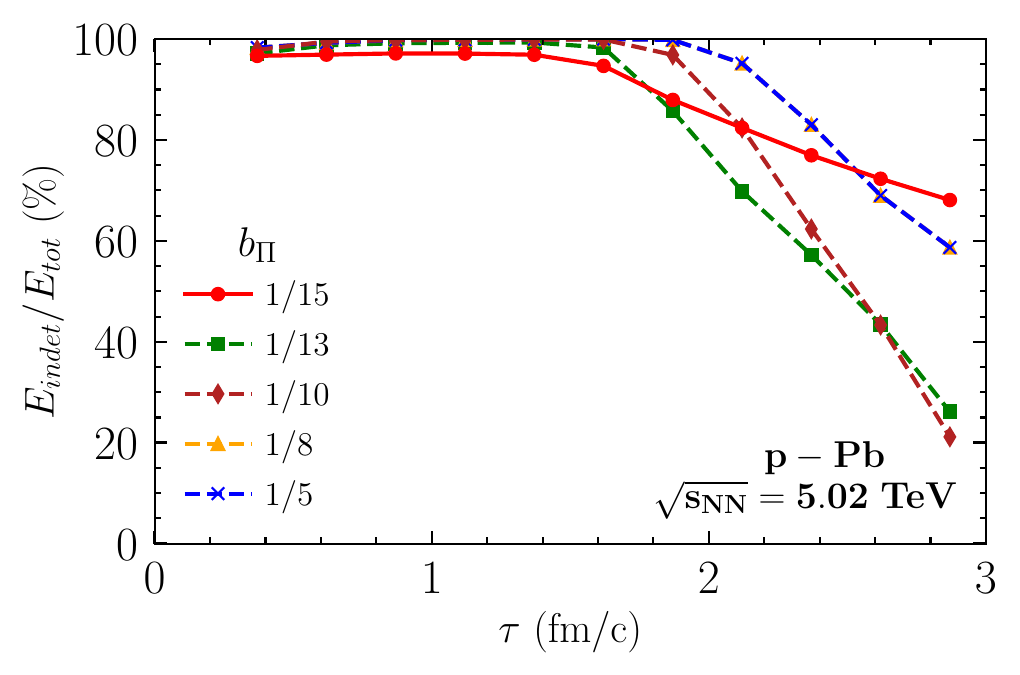}
    \caption{Different values for the parameter $b_\Pi$ and how it modifies the properties: acausality, causality, and indeterminate energy for p-Pb collision. A similar conclusion of Fig. \ref{fig13} can be obtained: larger $b_\Pi$ produces a reduction in acausality energy, but the conversion in causal energy is delayed.}
    \label{fig14}
\end{figure}

Increasing $b_\Pi$ results in a larger bulk relaxation time and diminishes the problematic term in equation $4f$. As seen in the first plot of Fig.\ \ref{fig13}, acausality is reduced when the bulk relaxation time is increased. We note, however, that a larger bulk relaxation time also means a slower evolution of the system, as shown in Fig.\ \ref{fig13}, where the blue curve, which corresponds to $b_\Pi = 1/3$, achieves only 98\% of causality at the end of hydrodynamics. A similar trend is observed for p-Pb in Fig.\ \ref{fig14}, where a larger $b_\Pi$ value reduces acausality but slows down the system’s evolution. Thus, there might exist an optimal range for $b_\Pi$ in which causality violations are greatly reduced or eliminated, and the system still can relax fast enough to ensure 100\% of causality within its lifetime. This optimal parameter might be determined from a Bayesian analysis by considering the causality violations in the selection of parameters. A better balance between the time to achieve sufficient conditions for causality and the initial weight of acausal cells could also potentially be found by changing the functional shape of the bulk-viscous relaxation time, besides adjusting its overall magnitude \cite{he2021interplaying}.

\begin{figure}[ht!]
    \centering
    \includegraphics[width=0.51\textwidth]{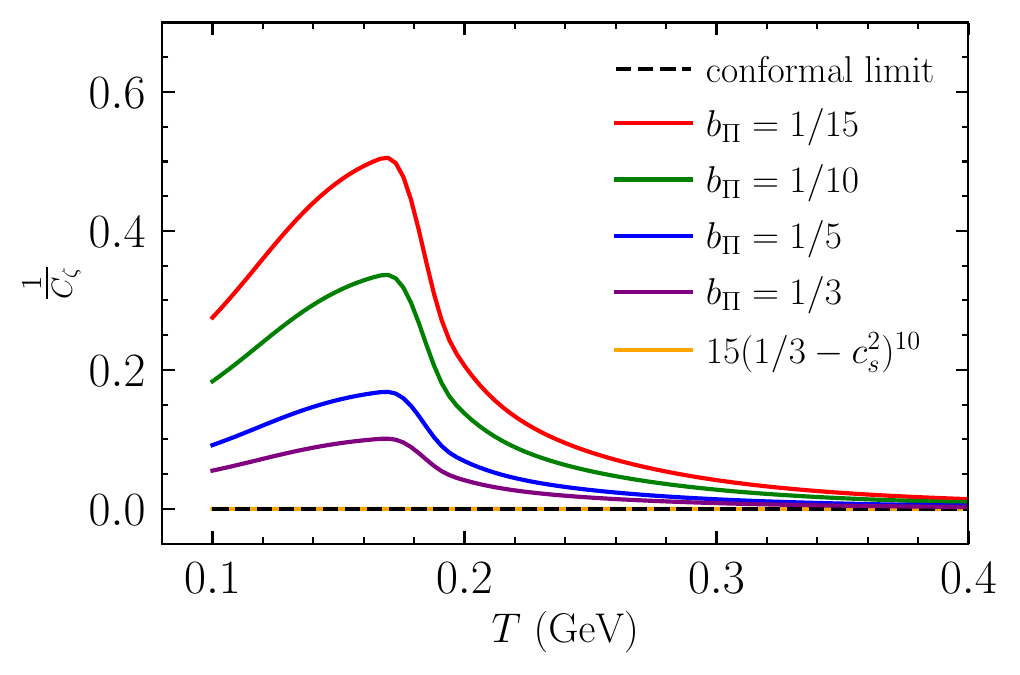}
    \caption{The dimensionless coefficient $1/C_\zeta$ for different values of $b_\Pi$ as a function of temperature.}
    \label{fig:inv_C_zeta_T}
\end{figure}

We also note the trend in both systems that acausality initially increases at early times before decreasing at later times. A possible explanation for this issue might come from the shape of the dimensionless coefficient $\frac{1}{C_{\zeta}} = \frac{\zeta}{\tau_{\Pi}(\epsilon + P)}$, which has the temperature dependence $\propto (1/3 - c_s^2)^{\alpha}$, with $\alpha=2$, and is plotted in Fig.\ \ref{fig:inv_C_zeta_T}. We also show, by plotting $\frac{1}{C_{\zeta}}$ for $\alpha = 10$, that it is possible to recover the conformal limit with a large enough value of $\alpha$.
Since this term is dominant even in the linear regime, as shown in Fig.~\ref{fig:1}, it can be used to illustrate the time evolution of causality violations. 

\begin{figure}
    \centering
    \includegraphics[scale=0.45]{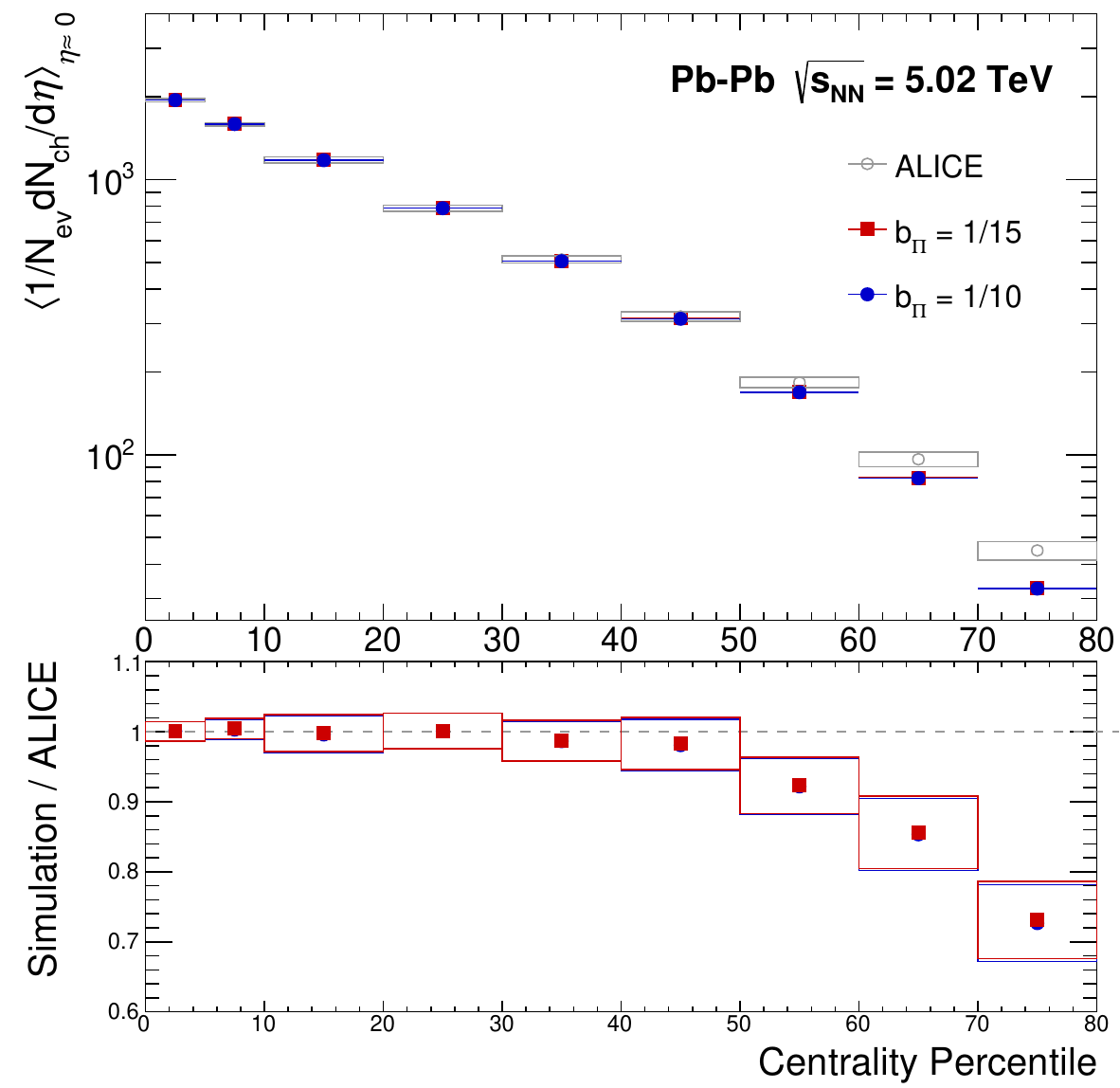} \\
    \includegraphics[scale=0.44]{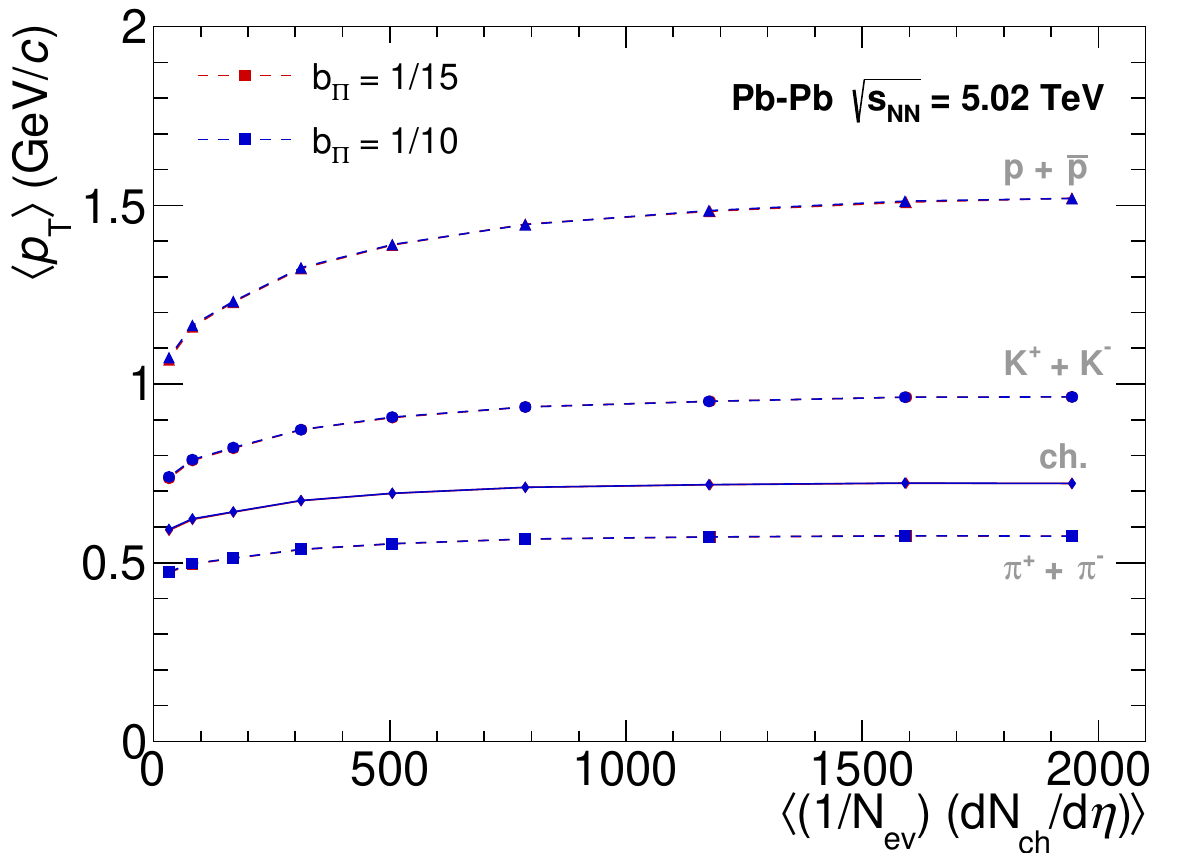} 
    \includegraphics[scale=0.44]{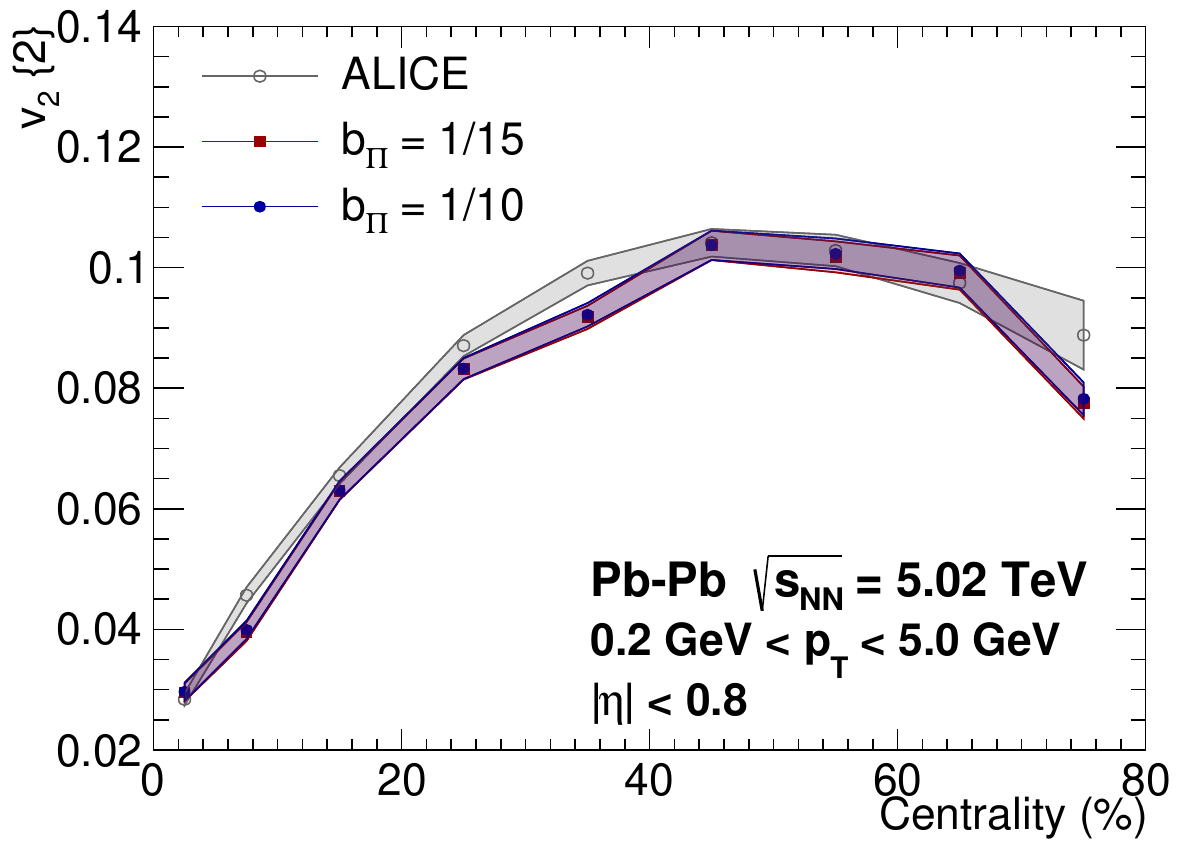}
    \caption{Multiplicity, mean transverse momentum, and elliptic flow coefficient for different values of the $b_{\Pi}$ parameter in Pb-Pb collisions. Experimental data comes from the articles \cite{adam2016centrality, adam2016anisotropic}. Changing the parameter does not affect these final-state observables determined by the simulation.}
    \label{fig15}
\end{figure}

\begin{figure}
    \centering
    \includegraphics[scale=0.45]{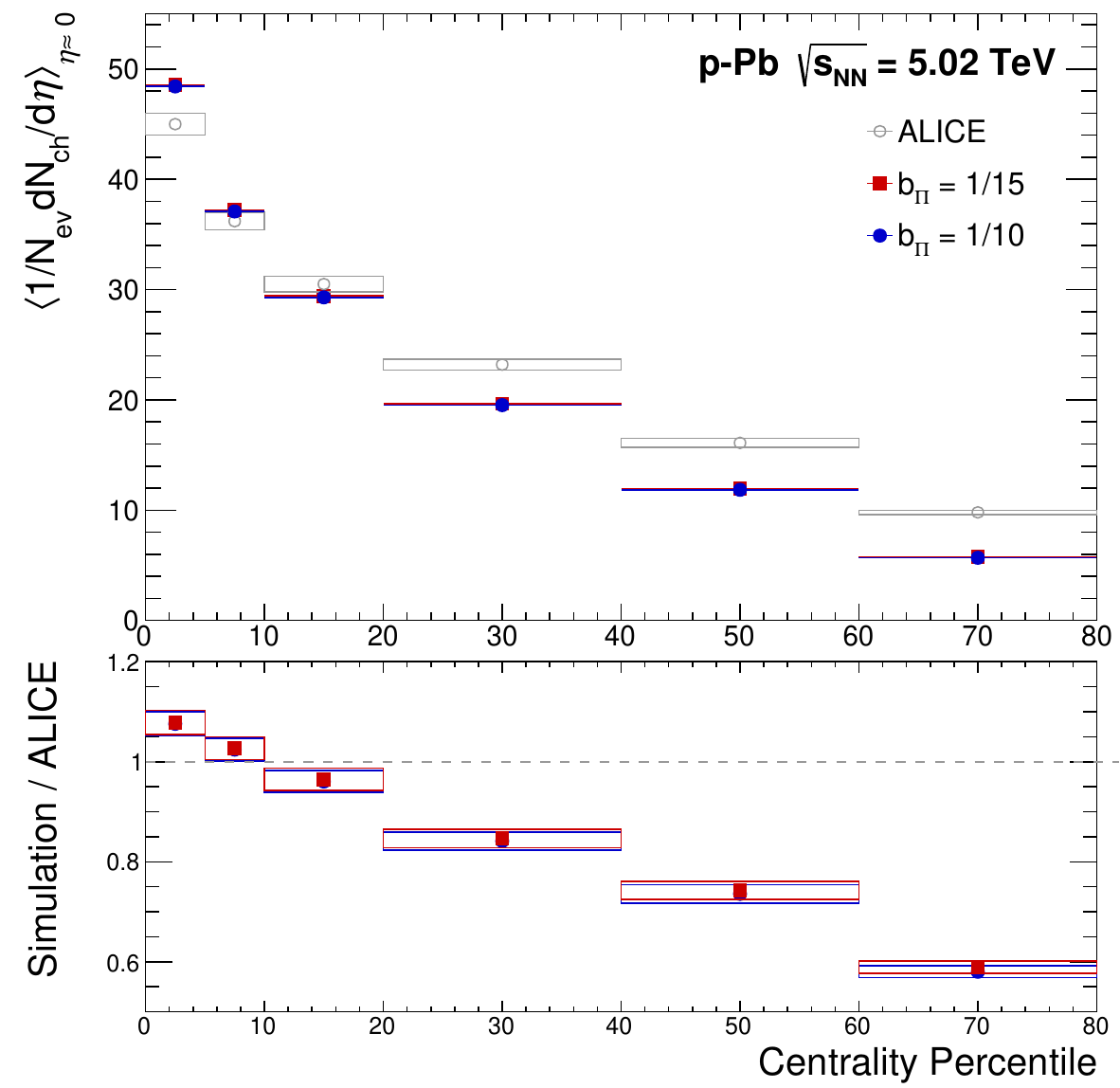} \\
    \includegraphics[scale=0.44]{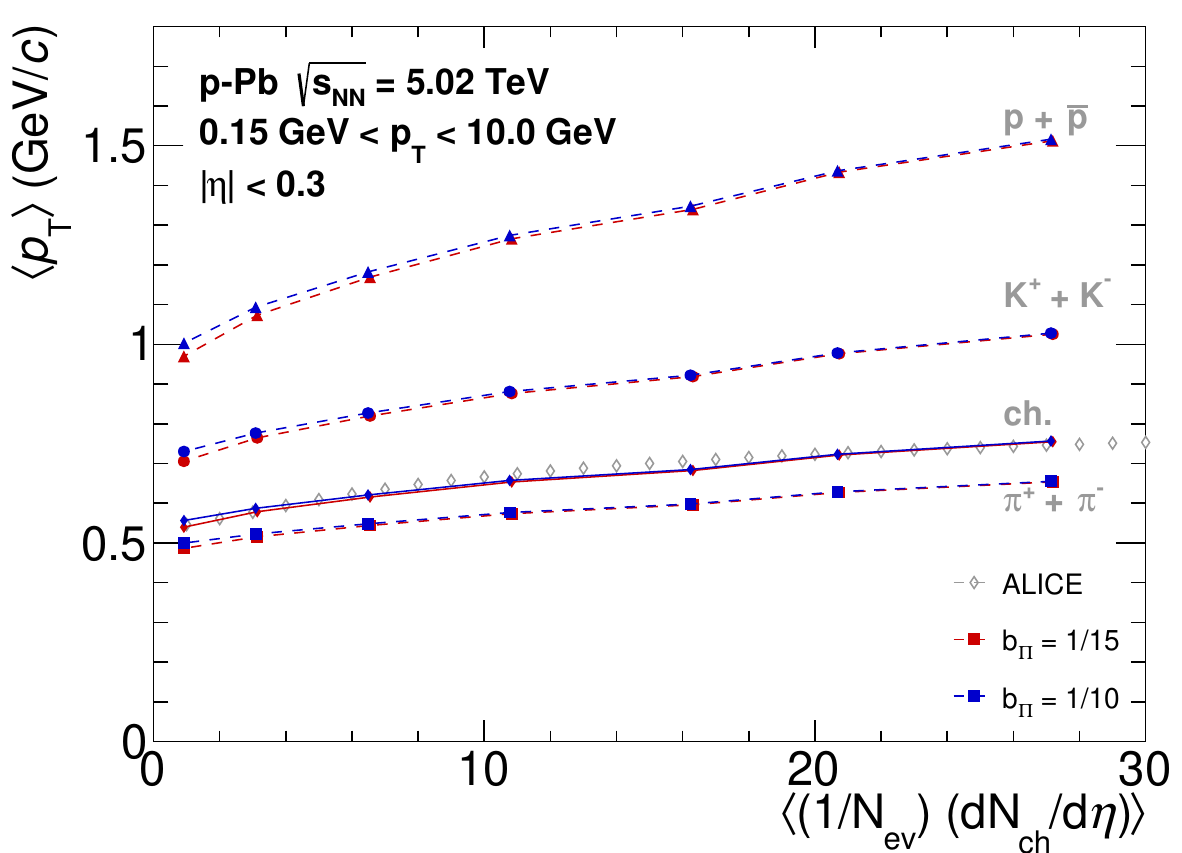}
    \includegraphics[scale=0.44]{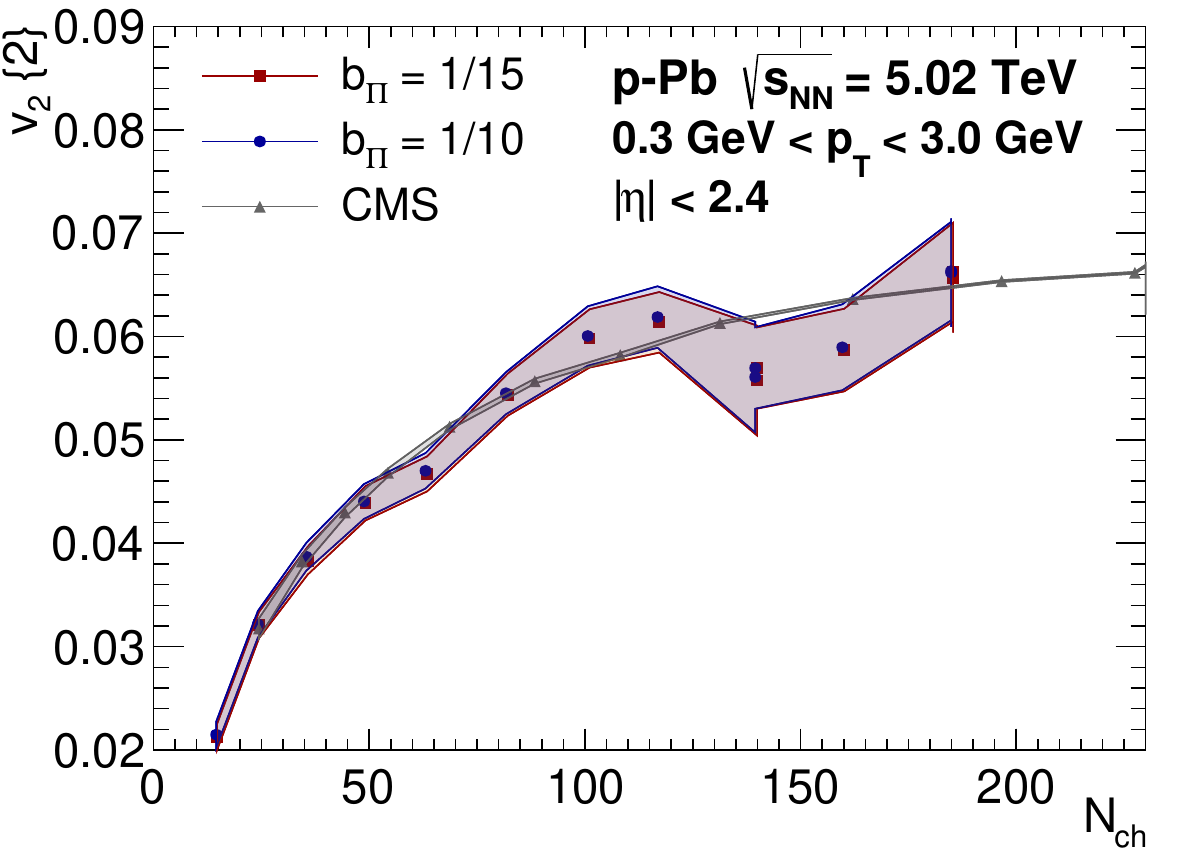}
    \caption{Multiplicity, transverse momentum, and elliptic flow coefficient for p-Pb collisions with different values for $b_\Pi$. Experimental data comes from the articles \cite{abelev2014multiplicity, abelev2013multiplicity}. Significant changes are not observed for different values of $b_\Pi$.}
    \label{fig16}
\end{figure}

An important question to consider is whether causality violation affects final state observables. If so, they might contaminate transport coefficients extracted in Bayesian studies. The findings above indicate that this effect can be reduced by varying the bulk relaxation time. We have probed three of those observables from our simulated events, namely the final state charged multiplicity, mean transverse momentum, and anisotropic flow coefficient $v_2$ for the standard value of $b_\Pi$ and also a different value which reduces the acausal behavior. We have employed our complete simulation chain in these simulations, including the particlization and hadronic cascade stages. We have not observed significant changes in these observables, in either of the systems, as can be seen in Fig.\ \ref{fig15} for Pb-Pb and in Fig.\ \ref{fig16} for p-Pb collisions. Thus, in this case, it is important that future Bayesian analyses consider the energy contained in acausal cells to quantify causality violations and their consequences.

\section{Conclusions}
\label{sec.conclusions}

In this work, we investigate the presence of acausal behavior in hydrodynamic simulations of heavy-ion collisions. This issue appears in large and small systems, with the latter showing a larger percentage of cells displaying causality violation (violations also increase as one moves towards ultraperipheral collisions, even in large systems). However, our results obtained using $\mathrm{T}_{\mathrm{R}} \mathrm{ENTo}$ initial conditions demonstrate that acausal cells carry a small part of the system's energy. Further work is needed to check if this statement remains true for other initial conditions where causality violations were more pervasive, such as the IP-Glasma initial conditions \cite{Gale:2012rq} investigated in \cite{plumberg2022causality}.  

Furthermore, we found that causality violations are more clearly related to the contributions coming from bulk viscosity. Additionally, we showed that acausality is significantly reduced when the velocity of free-streaming is modified to a value that reduces the effects of the bulk pressure. Finally, we observed that acausality was almost eliminated when we used a different parametrization for the bulk relaxation time. 

While increasing the overall magnitude of the bulk relaxation time drastically reduces the energy in acausal cells, it also increases the time for sufficient conditions for causality to be achieved. Further work is required to establish the optimal balance between these two effects, which may require changes to the temperature dependence of the relaxation time.

We hope our work contributes to drawing more attention to the issue of causality violations in hydrodynamic simulations, and stimulate future developments in this area. For example, it would be interesting to perform a Bayesian analysis that considers the acausality problem in its prior calibration and optimizes the causality in the system. Such a study could redefine the parametrization of the bulk relaxation time, which could then be applied in more general conditions. 

Finally, it would be important to find a way to have simulations with no acausal cells. Having simulations of relativistic viscous fluids where causality violation is found is, at best, disturbing. Such an issue must be fixed as one enters the so-called precision era of the hydrodynamic description of the quark-gluon plasma formed in heavy-ion collisions.    

\section*{Acknowledgments}
R.K. was funded by a CAPES Master's Fellowship. T.N.dS. acknowledges financial support from CNPq grant number 409029/2021-1. R.K. and T.N.dS acknowledge support from INCT-FNA research project 464898/2014-5. This research was funded by FAPESP grants number 2016/13803-2 (D.D.C.), 2021/04924-9 (A.V.G.), 2016/24029-6, 2018/24720-6 (M.L.), 2017/05685-2 (all) and 2018/07833-1 (M.H.). D.D.C., M.L., G.S.D., and J.T. thank CNPq for their financial support.  M.H. was partly supported by the National Science Foundation (NSF) within the framework of the MUSES collaboration, under grant number OAC-2103680. M.N.F. is supported by the Spanish MICINN grant PID2020-113334GB-I00 and by the Generalitat Valenciana through contract \mbox{CIAPOS/2021/74}. G.S.D. acknowledges financial support from Funda\c c\~ao Carlos Chagas Filho de Amparo \`a Pesquisa do Estado do Rio de Janeiro (FAPERJ), grant number E-26/202.747/2018. J.N. is partially supported by the U.S. Department of Energy, Office of Science, Office for Nuclear Physics under Award No. DE-SC0021301 and DE-SC0023861.

\bibliography{ref.bib}

\end{document}